\documentclass[
reprint,
superscriptaddress,
 amsmath,amssymb,
 prx
]{revtex4-2}

\usepackage{xcolor}
\usepackage{graphicx}
\usepackage{bm}% bold math
\usepackage[colorlinks=true,urlcolor=blue,citecolor=blue,linkcolor=blue]{hyperref}

\newcommand{\esp}[1]{\mathrm{e}^{#1}}
\newcommand{\contrast}{\tilde{C}}

\begin{document}

%%%%%%%%%%%%%%%%%%%%%%%%%%%%%%%%%%%%%%%%%%%%%%%%%%%%%%%%%%%%%%%%%%%%%%%%%%%%%%%%%%%%%%%%%%%%%%%%%%%%%%%%%%%%%%%5 TITLE
%\preprint{JOURNAL}
\title{Dimensional crossover in the superfluid-supersolid quantum phase transition
}
%\thanks{This is a draft}%

%%%%%%%%%%%%%%%%%%%%%%%%%%%%%%%%%%%%%%%%%%%%%%%%%%%%%%%%%%%%%%%%%%%%%%%%%%%%%%%%%%%%%%%%%%%%%%%%%%%%%%%%%%%%%%%5 AUTHORS
\author{Giulio Biagioni}
\thanks{These authors contribute equally}
\affiliation{CNR-INO, Sede di Pisa, 56124 Pisa, Italy}
\affiliation{Department of Physics and Astronomy, University of Florence, 50019 Sesto Fiorentino, Italy}

\author{Nicol\`{o} Antolini}
\thanks{These authors contribute equally}
\affiliation{CNR-INO, Sede di Pisa, 56124 Pisa, Italy}
\affiliation{LENS, University of Florence, 50019 Sesto Fiorentino, Italy}

\author{Aitor Ala\~na}
\affiliation{Department of Physics, University of the Basque Country UPV/EHU, 48080 Bilbao, Spain}

\author{Michele Modugno}
\affiliation{Department of Physics, University of the Basque Country UPV/EHU, 48080 Bilbao, Spain}
\affiliation{IKERBASQUE, Basque Foundation for Science, 48013 Bilbao, Spain}

\author{Andrea Fioretti}
\affiliation{CNR-INO, Sede di Pisa, 56124 Pisa, Italy}

\author{Carlo Gabbanini}
\affiliation{CNR-INO, Sede di Pisa, 56124 Pisa, Italy}

\author{Luca Tanzi}
\affiliation{CNR-INO, Sede di Pisa, 56124 Pisa, Italy}

\author{Giovanni Modugno}
\affiliation{CNR-INO, Sede di Pisa, 56124 Pisa, Italy}
\affiliation{Department of Physics and Astronomy, University of Florence, 50019 Sesto Fiorentino, Italy}
\affiliation{LENS, University of Florence, 50019 Sesto Fiorentino, Italy}

%%%%%%%%%%%%%%%%%%%%%%%%%%%%%%%%%%%%%%%%%%%%%%%%%%%%%%%%%%%%%%%%%%%%%%%%%%%%%%%%%%%%%%%%%%%%%%%%%%%%%%%%%%%%%%%5 DATE
%\date{\today}

%%%%%%%%%%%%%%%%%%%%%%%%%%%%%%%%%%%%%%%%%%%%%%%%%%%%%%%%%%%%%%%%%%%%%%%%%%%%%%%%%%%%%%%%%%%%%%%%%%%%%%%%%%%%%%%5 ABSTRACT
\begin{abstract}
\noindent We assess experimentally and theoretically the character of the superfluid-supersolid quantum phase transition recently discovered in trapped dipolar quantum gases. We find that one-row supersolids can have already two types of  phase transitions, discontinuous and continuous, that are reminiscent of the first- and second-order transitions predicted in the thermodynamic limit in 2D and 1D, respectively. The smooth crossover between the two regimes is peculiar to supersolids and can be controlled via the transverse confinement and the atom number. We justify our observations on the general ground of the Landau theory of phase transitions. The quasi-adiabatic crossing of a continuous phase transition opens new directions of investigation for supersolids.
\end{abstract}

%%%%%%%%%%%%%%%%%%%%%%%%%%%%%%%%%%%%%%%%%%%%%%%%%%%%%%%%%%%%%%%%%%%%%%%%%%%%%%%%%%%%%%%%%%%%%%%%%%%%%%%%%%%%%%%5 KEYWORDS

%%%%%%%%%%%%%%%%%%%%%%%%%%%%%%%%%%%%%%%%%%%%%%%%%%%%%%%%%%%%%%%%%%%%%%%%%%%%%%%%%%%%%%%%%%%%%%%%%%%%%%%%%%%%%%%5 MAKETITLE&TOC
\maketitle
%\tableofcontents

% check the REVTeX guide here -> https://d22izw7byeupn1.cloudfront.net/files/revtex/apsguide4-1.pdf
%%%%%%%%%%%%%%%%%%%%%%%%%%%%%%%%%%%%%%%%%%%%%%%%%%%%%%%%%%%%%%%%%%%%%%%%%%%%%%%%%%%%%%%%%%%%%%%%%%%%%%%%%%%%%%%5 MAIN
\section{\label{sec:intro}Introduction}

Supersolids are a fundamental phase of matter that mixes the properties of superfluids and crystals. Proposed more than 50 years ago \cite{Gr57,An69,Ch70,Le70}, a supersolid phase was recently observed in Bose-Einstein condensates of strongly magnetic atoms \cite{Ta19,Bo19,Ch19}, featuring simultaneous breaking of the global $U(1)$ and translational symmetries \cite{Ta19b,Gu19,Na19} and reduced moment of inertia under rotations \cite{Ta21}. The so-called dipolar supersolid requires a confinement along at least one spatial direction, so the lattice structure can develop in 1D \cite{Ta19,Bo19,Ch19} or in 2D \cite{No21}. The nature of the atomic interactions is such that the supersolid is of the cluster type, i.e. each lattice site hosts many atoms, of the order of one thousand. This realizes the scenario first depicted by E. Gross \cite{Gr57}, ensuring strong superfluidity effects. Thanks to the tunability of the interactions, it is possible to study the quantum phase transition between the superfluid phase, a standard Bose-Einstein condensate, and the supersolid phase \cite{Ta19,Bo19,Ch19,He21}, in addition to the classical phase transition from a thermal gas to a supersolid \cite{So21}.

Here we focus on the superfluid-supersolid quantum phase transition, a new fundamental phase transition whose character has not yet been assessed. In the thermodynamic limit, general models based on soft-core interactions predict first-order transitions in 2D \cite{Po94,Ma13}, and second order transitions in 1D \cite{Se08}, in analogy to standard crystallization transitions. For dipolar supersolids, the predicted scenario is more complex, with two types of first-order transitions for 2D lattices \cite{Zh19,Zh21} and both first- and second-order transitions for 1D lattices \cite{Ro19,Bl20} depending on the density. In the experiments, the problem is complicated not only by the dimensionality varying continuously between 1D and 2D but also by the finite size and the inhomogeneity due to the presence of harmonic potentials. So far, the quantum phase transition was crossed only for supersolids with 1D lattice structures. Some of the studies reported partial indications of a discontinuous \cite{Ta19,Bo19,Ta19b} or a continuous transition \cite{Pe21}, while others did not address the character of the transition \cite{He21}. So, it is not clear whether in  one-row dipolar supersolids one can observe continuous, discontinuous or both types of quantum phase transitions, nor how these transitions relate to the general theory. For trapped supersolids with 2D lattice structures, numerical simulations predict discontinuous transitions \cite{He21c,Bl21}.

Although there are materials featuring density-modulated cluster phases similar to the dipolar supersolid, such as the smectic phase of liquid crystals \cite{Ch95} or the Rosensweig phase of ferrofluids \cite{ro13,Ch95}, these phases are normally reached via classical transitions and at fixed dimensionality. Also the pair-density-wave quantum phases recently discovered in He superfluids \cite{Le19,Sh20} and in superconductors \cite{Ag20}, were so far only studied in 2D geometries. So, no results from existing phase transitions can be employed to draw predictions about supersolids.

In this work we perform a combined experimental and theoretical investigation to assess the character of the superfluid-supersolid phase transition for one-row dipolar supersolids. We find that the phase transition changes smoothly from continuous to discontinuous as the lattice geometry is changed from 1D towards 2D, in analogy to the second-order and first-order crystallization transitions in 1D and 2D, respectively. There is, however, an important novelty due to the very nature of supersolids: the discontinuous transition can persist also for apparently 1D lattices, due to the intrinsic density background of the supersolid which keeps a 2D structure. In addition, the cluster nature of the dipolar supersolid limits the impact of finite-size effects and allows to observe the character of the phase transition also in systems with very short lattices.

Our analysis reconciles previous results, justifies the observations on the general grounds of the Landau theory, and shows how to achieve in a controllable way continuous or discontinuous quantum phase transitions in one-row supersolids. Continuous transitions are particularly interesting as they allow to realize excitation-free supersolids, opening new directions of investigation.

\section{\label{sec:Landau}Formation of the dipolar supersolid in an infinite system}

\begin{figure}
    \centering
    \includegraphics[width=0.48\textwidth]{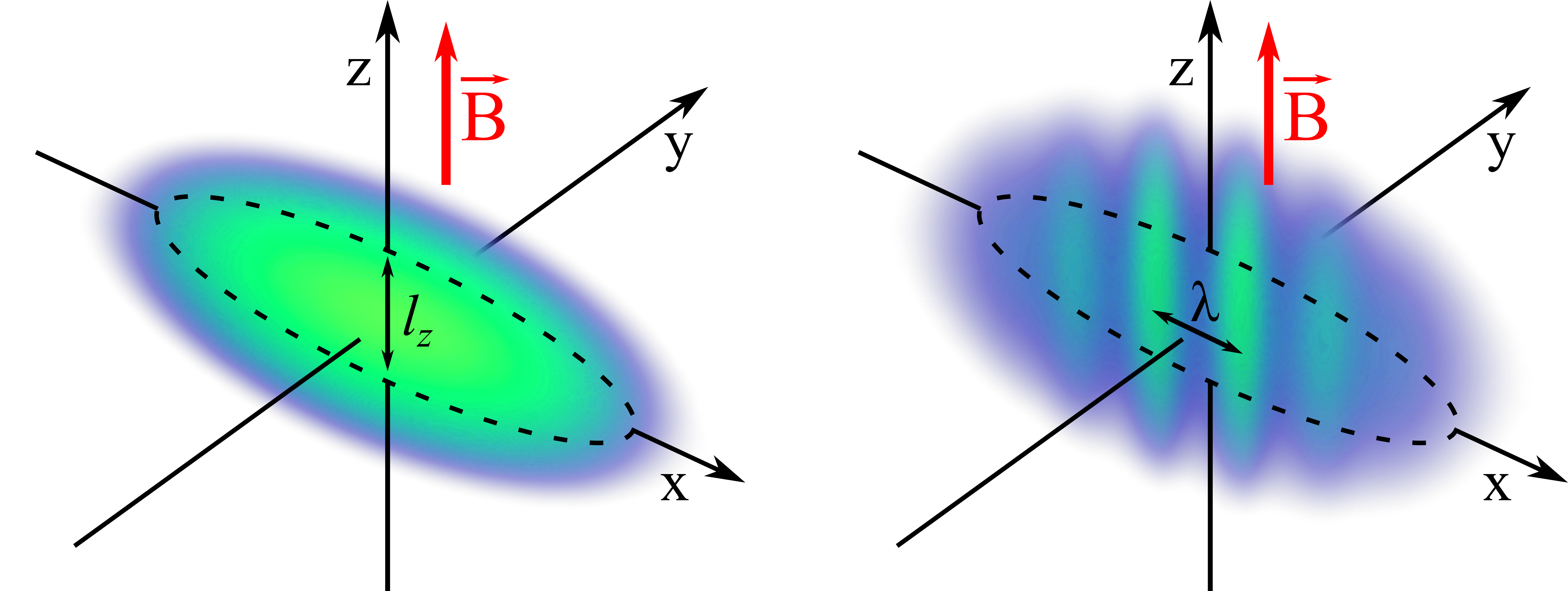}
    \caption{Geometry of the elongated superfluid (left) and the one-row supersolid (right) commonly realized in experiments. The atomic magnetic dipoles are aligned in the $z$ direction by the magnetic field $B$. The relevant lengthscales are the harmonic confinement length  $\ell_z$ and the supersolid lattice spacing $\lambda$.}
    \label{fig:SS-SF}
\end{figure}

The geometry of the dipolar supersolids investigated experimentally is sketched in Fig. \ref{fig:SS-SF}. The magnetic dipoles are aligned in the $z$ direction by a magnetic field $B$, and an anisotropic harmonic potential is present in the three spatial directions.

To get intuition into the physics at play, we start from the infinite case, removing the confinement in the $x-y$ plane. We consider a dipolar quantum gas at $T=0$, described by the wavefunction $\psi$, with density $\rho = |\psi|^2$. The energy of the system is
(see Appendix~\ref{sec:methods_num})
\begin{equation}\label{eq:TotalEnergy}
    E[\psi] = E_{kin} + E_{trap} + E_{cont} + E_{dd} + E_{LHY}.
\end{equation}
The first term is the kinetic energy, $E_{kin}\propto|\nabla \psi|^2$. $E_{trap}$ describes a vertical harmonic confinement with frequency $\nu_z$. The mean-field contact and dipolar interactions scale as $E_{cont} \propto a_s \rho^2$ and $E_{dd}\propto a_{dd}\rho^2$, where $a_s$ is the s-wave scattering length and $a_{dd}= {\mu_0\mu^2 m}/({12 \pi \hbar^2})$ is the dipolar length associated with particles with magnetic dipole $\mu$ and mass $m$. The last term is the so-called Lee-Huang-Yang (LHY) energy, which describes the zero-point energy of quantum fluctuations in the local-density approximation \cite{Li12}, and scales as $E_{LHY}\propto \rho^{5/2}$.

The transition from the superfluid to the supersolid is crossed by reducing the repulsive scattering length $a_s$, thus increasing the relative strength of the dipolar interaction. In the supersolid phase, the dipolar energy is reduced due to the enhancement of head-to-tail arrangement of the dipoles within each cluster, but there is an increase of both contact energy, coming from an increase in the peak density, and in kinetic energy, due to the density modulation. When the dipolar gain is larger than the contact and kinetic costs, the transition takes place.

The lattice period $\lambda$ of the supersolid is of the order of the harmonic length in the direction of the $B$ field, $\ell_{z}=\sqrt{\hbar/2\pi m \nu_z}$, as sketched in Fig.~\ref{fig:SS-SF}, not far from the wavelength of the roton excitation mode of the superfluid \cite{Sa03,Ch18}. Hereinafter, we will assume to freeze out the $z$ direction and we will refer to 1D and 2D supersolids depending on the breaking of translational invariance occurring along one or two directions in the $x-y$ plane.

At the mean-field level the supersolid is unstable, since the dipolar energy becomes more and more negative for increasing density. Collapse is eventually prevented by the repulsive LHY energy term. By further decreasing the scattering length, the system crosses a second phase transition towards a droplet crystal, where the superfluid background disappears and coherence between clusters is lost \cite{Ta19,Bo19,Ch19}. In the present work we  do not study the latter transition.

Generally, the main features of a phase transition are captured by the Landau theory, in which the ground state of the system is determined by the behavior of the free energy as a function of the order parameter. For crystallization phase transitions, the typical order parameter is the contrast \textit{C} of the density modulation, which is zero in liquid-like phases and different from zero in crystal-like phases. The energy difference between the state with $C=0$ and that with $C \neq 0$ can be expanded in powers of \textit{C} as
\begin{equation}
    \Delta E~\simeq
    ~a~C+b~C^2+c~C^3+d~C^4, %+...
\label{eq:LandauExpansion}
\end{equation}
where the values of the coefficients $a,b,c,d$ determine the character of the phase transition. In the infinite case under consideration, the linear term $a$ is always zero, but it appears when a trap is present, see Appendix \ref{app:landau_model}.

As usual in the context of the Landau theory, the symmetries of the ground state provide information on the character of the phase transition.
Close to the transition, where $C\ll 1$, a sinusoidal modulation is a good approximation for the supersolid ground state density \cite{Po94,Zh19,Bl20}:
\begin{equation}\label{eq:Density}
    \rho (\bm{r})=\rho_0 \big[1+ C\sum_i \cos ( \bm{k}_i\cdot \bm{r}) \big],
\end{equation}
where $\rho_0$ is the average density and $\bm{k}_i$ the lattice wave-vectors defining the lattice dimensionality.

\begin{figure}
    \centering
    \includegraphics[width=0.48\textwidth]{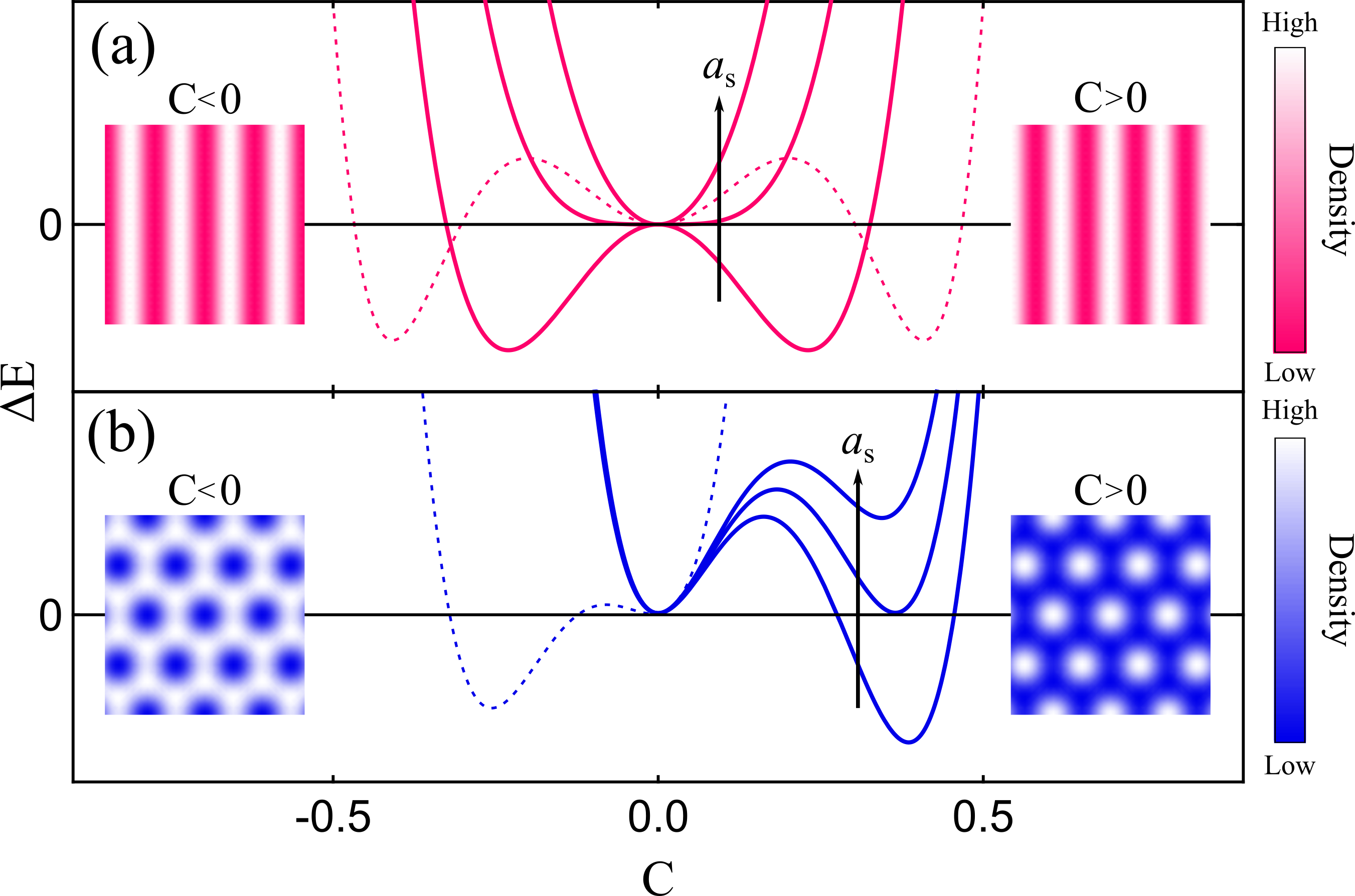}
    \caption{Landau theory of the superfluid-supersolid quantum phase transition in the thermodynamic limit. Scenarios for second-order phase transitions in 1D (a) and first-order phase transitions in 2D (b). Solid lines are the typical behavior of the free energy as a function of the order parameter $C$ for varying scattering length. Dashed lines are examples of the free energy in the LHY-dominated regime. Insets show the lattice structure.
    }
    \label{fig:Fig1}
\end{figure}

In 1D the supersolid lattice is characterized by a single wave vector $\bm{k}$. In this case, ansatz~(\ref{eq:Density}) is symmetrical with respect to the substitution $C \rightarrow -C$, which leads only to an overall displacement of the lattice structure. Therefore, states which differ only in the sign of the contrast are physically equivalent. The free energy  Eq.~(\ref{eq:LandauExpansion}) must be an even function of $C$, i.e. $c_{1D}=0$. The transition occurs when, by lowering the scattering length, the dipolar energy overcomes the contact energy and reverses the sign of the quadratic coefficient $b_{1D}$. Therefore, the transition is of the second order, see Fig.~\ref{fig:Fig1}(a).

The 2D case was studied in Refs.~\cite{Po94,Zh19}. The supersolid lattice is triangular, which is the closest packing configuration, with lattice wave-vectors of equal length satisfying $\bm{k}_1+\bm{k}_2+\bm{k}_3=0$. In this case, the previous symmetry is lost, as ansatz~(\ref{eq:Density}) represents two very different states depending on the sign of $C$. For $C>0$ it corresponds to a triangular lattice of density maxima, for $C<0$ to a triangular lattice of holes, i.e. a honeycomb lattice. Therefore, the free energy expansion Eq.~(\ref{eq:LandauExpansion}) can contain odd terms, in particular the cubic term $c_{2D}$. %\neq 0$.
This results in a discontinuous phase transition, characterized by the presence of a  metastable state with finite contrast and a jump in the order parameter. In dipolar supersolids, the sign of the cubic term $c_{2D}$ depends on the competition between kinetic and LHY energy \cite{Zh19}. At relatively low density, kinetic energy dominates and leads to $c_{2D}<0$, resulting in a phase transition towards a triangular lattice with $C>0$, see solid lines in Fig.~\ref{fig:Fig1}(b). At high densities, the LHY contribution becomes important and can reverse the sign of $c_{2D}$. In this case, the ground state is a honeycomb lattice with $C<0$, see the dotted line of Fig.~\ref{fig:Fig1}(b).

We note that the Landau theory allows for a discontinuous phase transition also for the 1D case, when the free energy is an even function of $C$. This happens if the quartic coefficient $d_{1D}$ is negative \cite{Ch95}. To ensure stability, one has to expand the free energy to the sixth order. By decreasing the scattering length, one crosses a discontinuous transition as shown by the dashed line in Fig. \ref{fig:Fig1}(a). For the dipolar supersolid, $d_{1D}$ is determined by the competition between kinetic energy (positive contribution) and LHY energy (negative contribution). At very high densities, therefore, when the LHY energy dominates,we expect a first-order transition also in 1D. Since the LHY term is the zero-point energy of quantum fluctuations, the discontinuous transition in this regime belongs to the class of fluctuation-induced first-order phase transitions, as those found in some types of superconductors and liquid crystals \cite{Ha74}. Such an effect can explain the numerical observations in Ref. \cite{Bl20}.

\section{\label{sec:numerical} Theoretical phase diagram in the trapped system in equilibrium}

Moving to the finite-size systems studied in experiments, one should replace the thermodynamic concepts of first- and second-order phase transitions with those of discontinuous and continuous transitions. In addition, the inhomogeneity of the density in the harmonic potentials generally leads to a coexistence of the two phases.

In the supersolids realized so far, the harmonic confinement in the $x-y$ plane is typically anisotropic, leading to the formation of one-row lattices. One would naturally associate such configuration with the second-order phase transitions of the infinite system. However, various experiments \cite{Ta19,Bo19,Ta19b} and numerical simulations \cite{Ro19,Bl20} have shown the presence of apparently discontinuous transitions. To clarify the scenario, in this section we present a detailed analysis of the equilibrium states of trapped supersolids, obtained via numerical simulations.

To characterize the typical trapped systems, we fixed two of the trap frequencies to realistic values, $\nu_x$=20~Hz and $\nu_z$=80~Hz, which result in a supersolid elongated in the $x$ direction. We then varied both the transverse frequency $\nu_y$ and the atom number $N$ in realistic ranges, and for each set of values we studied the evolution of the ground state of the system as a function of the scattering length $a_s$. The ground state was obtained by minimizing numerically the energy in Eq.~(\ref{eq:TotalEnergy}), with the addition of trap energies also in the $x$ and $y$ directions.

\begin{figure}
    \centering
    \includegraphics[width=0.48\textwidth]{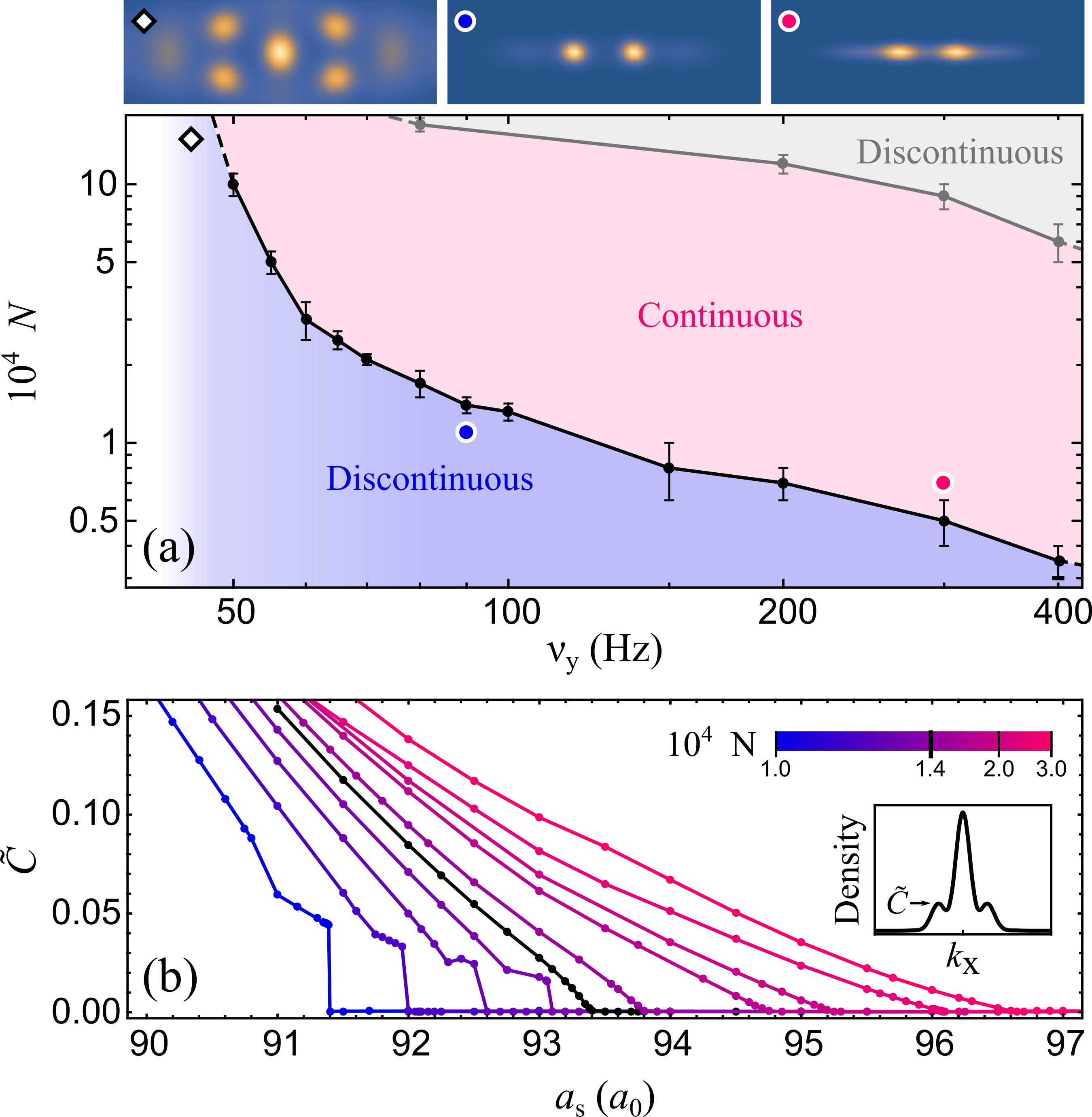}
    \caption{Character of the superfluid-supersolid phase transition from numerical simulations. (a) Phase diagram in the $N - \nu_y$ plane. Black dots correspond to the boundary between the continuous and discontinuous regimes. Gray dots mark the onset of the LHY-dominated regime. The three samples of the supersolid density in the upper panels, from right to left, correspond to: one-row continuous (magenta circle), one-row discontinuous (blue circle), and  two-rows (white diamond). (b) Contrast as a function of the scattering length along the crossover, for different values of $N$, and $\nu_y=90$~Hz. The curve corresponding to the center of the crossover is plotted in black. Solid lines are a guide for the eye. Inset: typical momentum distribution in the supersolid phase.}
    \label{fig:PhaseDiagram}
\end{figure}

Figure~\ref{fig:PhaseDiagram} presents a summary of the simulations. Fig.~\ref{fig:PhaseDiagram}(a) shows the presence of various regions of continuous and discontinuous transitions in the $N - \nu_y$ plane. For small $\nu_y$, the trap approaches a circular shape in the $x-y$ plane and the supersolid forms on two rows with a triangular structure. There, we find a discontinuous transition, in analogy with the first-order phase transition in 2D discussed in Section~\ref{sec:Landau}. By increasing the transverse frequency, we continue to observe discontinuous transitions although the supersolid apparently forms on a single row. The character of the transition changes gradually from discontinuous to continuous, reaching the 1D scenario of the previous section only for large $\nu_y$. This results in a smooth crossover from discontinuous (2D-like) transitions for small $N$ and $\nu_y$ towards continuous (1D-like) transitions for large $N$ and $\nu_y$. As we will show in the rest of this section, the shape of such 2D-1D crossover is related to the very nature of the dipolar supersolid. For large $N$, Fig.~\ref{fig:PhaseDiagram}(a) shows a second regime of discontinuous transitions, which we will discuss later. The regime of small $N$ is instead irrelevant, since the supersolid lattice shrinks to a single cluster.

The specific order parameter we consider is the contrast in momentum space $\contrast$, i.e. the height of the Fourier peak at $k=2\pi/\lambda$, which in the limit of small contrasts is related to the real space one by $\contrast=C^2/16$ (see Appendix~\ref{sec:methods_num}). As we will see in the next section, this choice is motivated by the presence of a related experimental observable.  Fig~\ref{fig:PhaseDiagram}(b) shows examples of the phase transitions for fixed $\nu_y$=90 Hz and variable $N$. While for large atom numbers $\contrast$ changes smoothly with $a_s$, for smaller $N$ the transition is discontinuous, with a finite jump in the contrast. We arbitrarily chose the center of the crossover as the smallest value of $N$ that gives a jump in $\contrast$ at the transition point smaller than 0.001.

As shown by the sample distributions in Fig~\ref{fig:PhaseDiagram}(a), the supersolid lattice forms in the central region of the system, while the low-density sides tend to remain in the superfluid regime. However, their weight is normally very small, so the behavior of the system is dominated by the supersolid.

\begin{figure}
    \centering
    \includegraphics[width=0.48\textwidth]{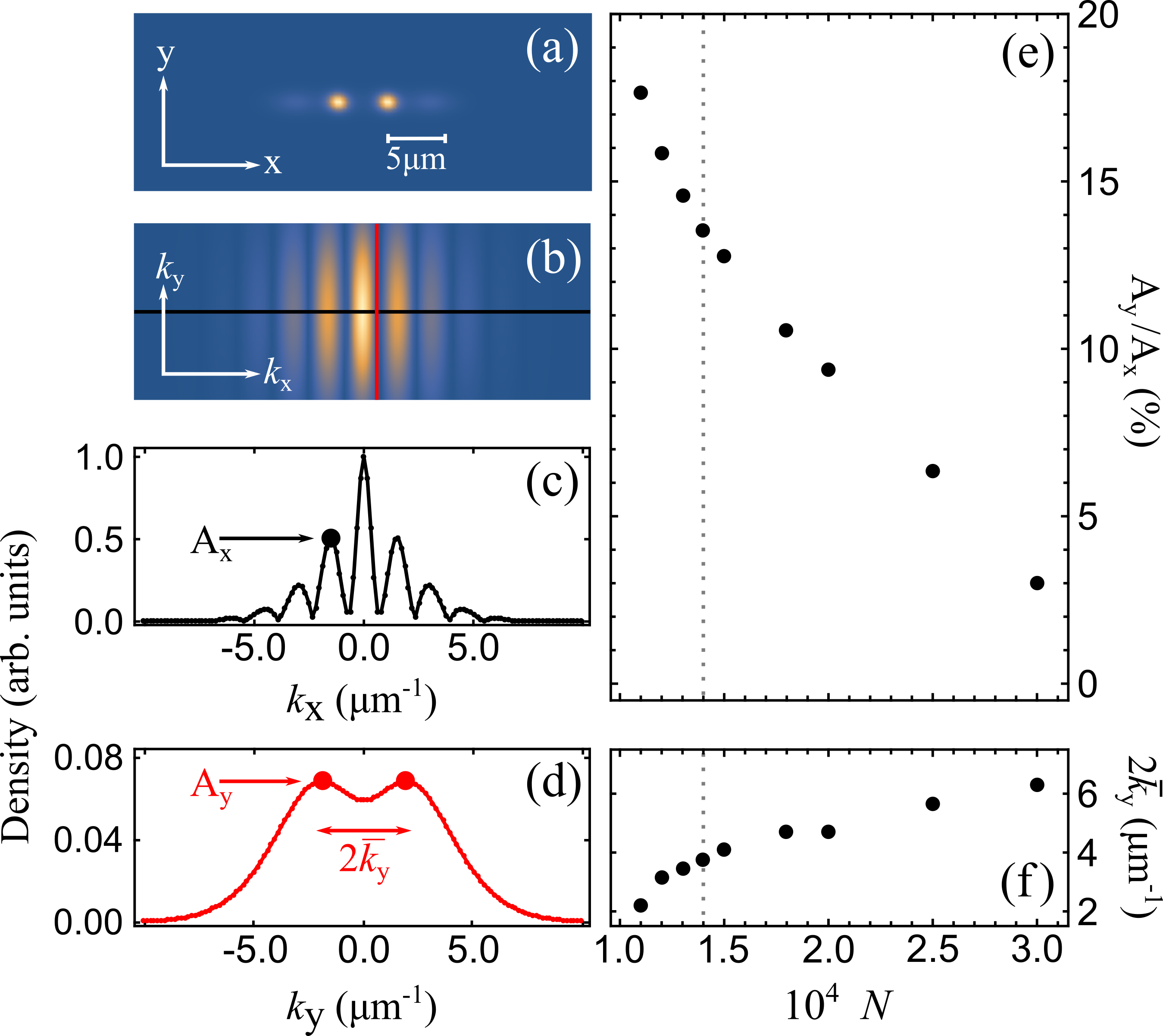}

    \caption{Two-dimensional structure of the density background across the dimensional crossover. Real-space (a) and Fourier- space (b) density of a typical simulated supersolid. Cuts in the Fourier space along $x$ (black line)  and $y$  (red line)  directions in (b), which are shown in panels (c) and (d) respectively, reveal the presence of peaks along both directions. In (e) ratio of the Fourier amplitudes along $y$ and $x$, and in (f) the Fourier spacing along $y$, for the simulations of Fig.~\ref{fig:PhaseDiagram}(b),  at $\contrast\simeq 0.04\%$. The vertical dotted lines mark the center of the crossover.}
    \label{fig:DimensionalOrigin}

\end{figure}

Although in most of the phase diagram the supersolid seems to develop with a single row of maxima, a Fourier analysis reveals the presence of a triangular structure of the density background. An example of such analysis is shown in Fig.~\ref{fig:DimensionalOrigin}. One notes the presence of Fourier peaks not only along the $x$ direction but also along the $y$ direction, although with much smaller amplitude and larger momentum, see Fig.~\ref{fig:DimensionalOrigin}(c-d).  So, even in the presence of a single row of principal density maxima and a tight transverse confinement, there is clearly a persistence of the triangular structure of the 2D supersolid. Although the amplitude of the triangular structure is small, it leads to discontinuous transitions. Both amplitude and momentum change continuously when moving from the discontinuous to the continuous side of the crossover, see Fig.~\ref{fig:DimensionalOrigin}(e-f). Interpreting the crossover in terms of the Landau theory, we conclude that the main effect of an increasing atom number is to suppress progressively the 2D triangular structure. This results in a progressive suppression of the cubic terms in the energy expansion, especially the kinetic energy term, which weakens the discontinuity until the transition becomes continuous. There is a similar effect increasing $\nu_y$ at constant $N$.

The prevalence of discontinuous transitions in the phase diagram of Fig.~\ref{fig:PhaseDiagram}(a) is due to a change of the compressibility of the supersolid for varying atom numbers. At small $N$, the clusters are more compressible and the triangular structure can be deformed more easily by transverse confinement. This favours discontinuous transitions over continuous ones.
Additional analysis of the simulated density distributions shows that the transverse size $\sigma_y$ of the supersolid at the phase transition at the center of the crossover is very close to the harmonic oscillator length $\ell_y$, i.e. the size of a non-interacting system, see Appendix~\ref{sec:datanum}. This supports the idea that the  transition becomes discontinuous when there is enough space in the trap to accommodate a 2D structure.

In the phase diagram of Fig.~\ref{fig:PhaseDiagram}(a) we observe also two clear effects of the LHY energy term in Eq.~(\ref{eq:TotalEnergy}), which is normally small until one reaches the regime of large densities, i.e. large $N$. The first effect is the presence of the additional continuous-discontinuous crossover for large $N$. This is due to a gradual increase of the LHY energy as the supersolid contrast increases, since the density also increases. The LHY contribution reverses the sign of the quartic term in the Landau expansion and changes the transition from continuous to discontinuous, as discussed in Section \ref{sec:Landau}. It is therefore a fluctuation-induced phase transition, related to the one found in the infinite case of Ref. \cite{Bl20}. The second effect is a gradual elongation along $y$ of the clusters for increasing $N$ and decreasing $\nu_y$. This effect is due to the increasing cost in LHY energy, which favors the formation of a 1D lattice of ``stripes'' directed along $y$ instead of a triangular lattice, as observed in previous simulations \cite{Zh21,He21b}. This effect broadens the region of continuous phase transitions in the upper left part of the $N - \nu_y$ plane. See Appendix \ref{sec:methods_num} for further discussion.

Finally, in the simulations we observe a small modulation appearing in discontinuous transitions just before the jump to higher contrast states, resulting in a mixing of continuous and discontinuous transitions. In the framework of the Landau model, we interpretate such an effect as a consequence of the trapping energy, which introduces a term linear in $C$ in the energy expansion (see Appendix \ref{app:landau_model}). This term is responsible for a small asymmetry between $C$ and $-C$, which tends to produce a weak modulation of the density even on the superfluid side. This effect was seen previously in numerical simulations of dipolar Bose-Einstein condensates \cite{Wilson:2008}.

\section{\label{sec:experimental} Experimental evidence of continuous and discontinuous phase transitions}

\begin{figure}
    \centering
    \includegraphics[width=0.48\textwidth]{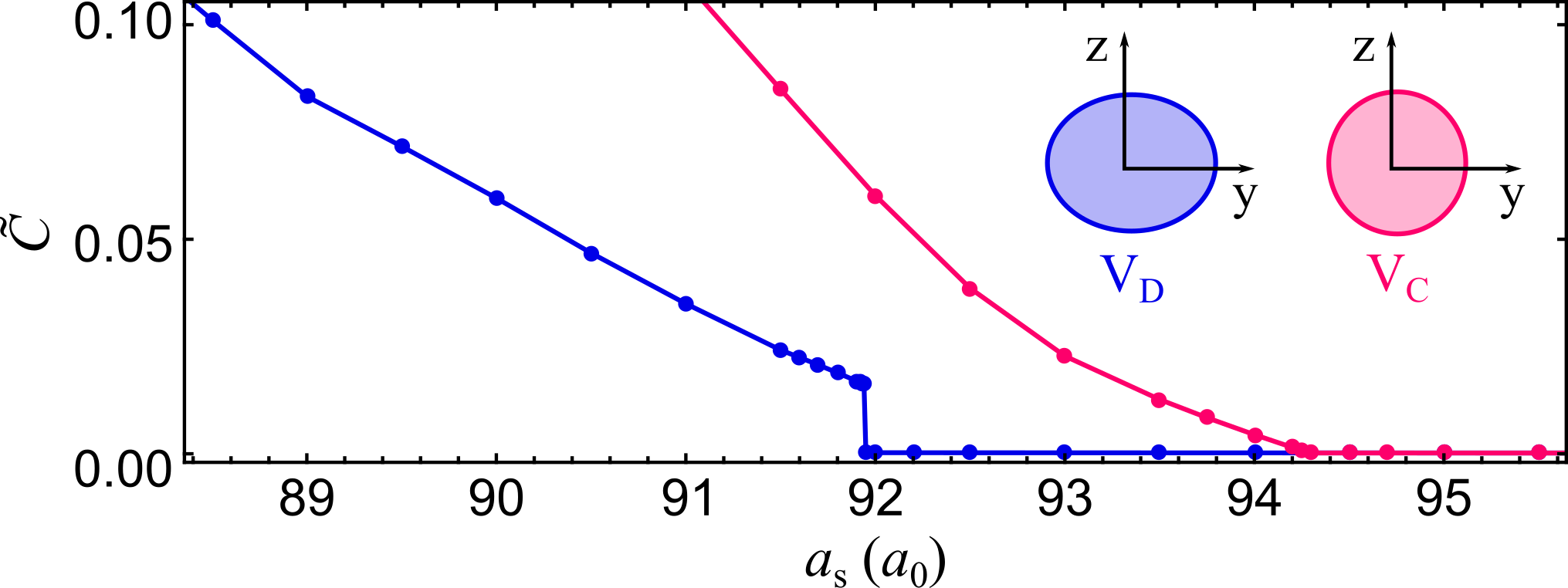}
    \caption{Equilibrium transitions in the experimental configurations. Simulated momentum-space contrast vs scattering length for harmonic potentials $V_D$ (blue) and and $V_C$ (magenta) with $N=3\times 10^4$. Inset: shape of the contour lines in the $y-z$ plane for the two potentials.}
    \label{fig:traps}
\end{figure}

We tested experimentally the theoretical predictions on a supersolid made of $^{162}$Dy magnetic atoms with dipolar length $a_{dd}=130~a_0$, trapped in optical potentials. To move from continuous to discontinuous transitions it is sufficient to change the aspect ratio of the harmonic potential in the $y-z$ plane, at constant atom number. We chose in particular two potentials on the two sides of the crossover. Potential $V_C$, with frequencies $(\nu_x, \nu_y, \nu_z)=(15.0~(0.7), 101.0~(0.3), 93.9~(0.6))$~Hz, provides a strong confinement along $y$, leading to a continuous transition. Potential $V_D$, with frequencies $(21.8~(1.0), 67.0~(0.8), 102.0~(0.7))$~Hz, provides instead a weaker confinement along $y$, leading to a discontinuous transition. Fig.~\ref{fig:traps} shows the predicted evolution of the contrast $\contrast$ with the scattering length from the ground-state simulations. In both cases, the mean atom number at the transition is about $N=3\times 10^4$. Both configurations correspond to one-row supersolids.

\begin{figure}
    \centering
    \includegraphics[width=0.48\textwidth]{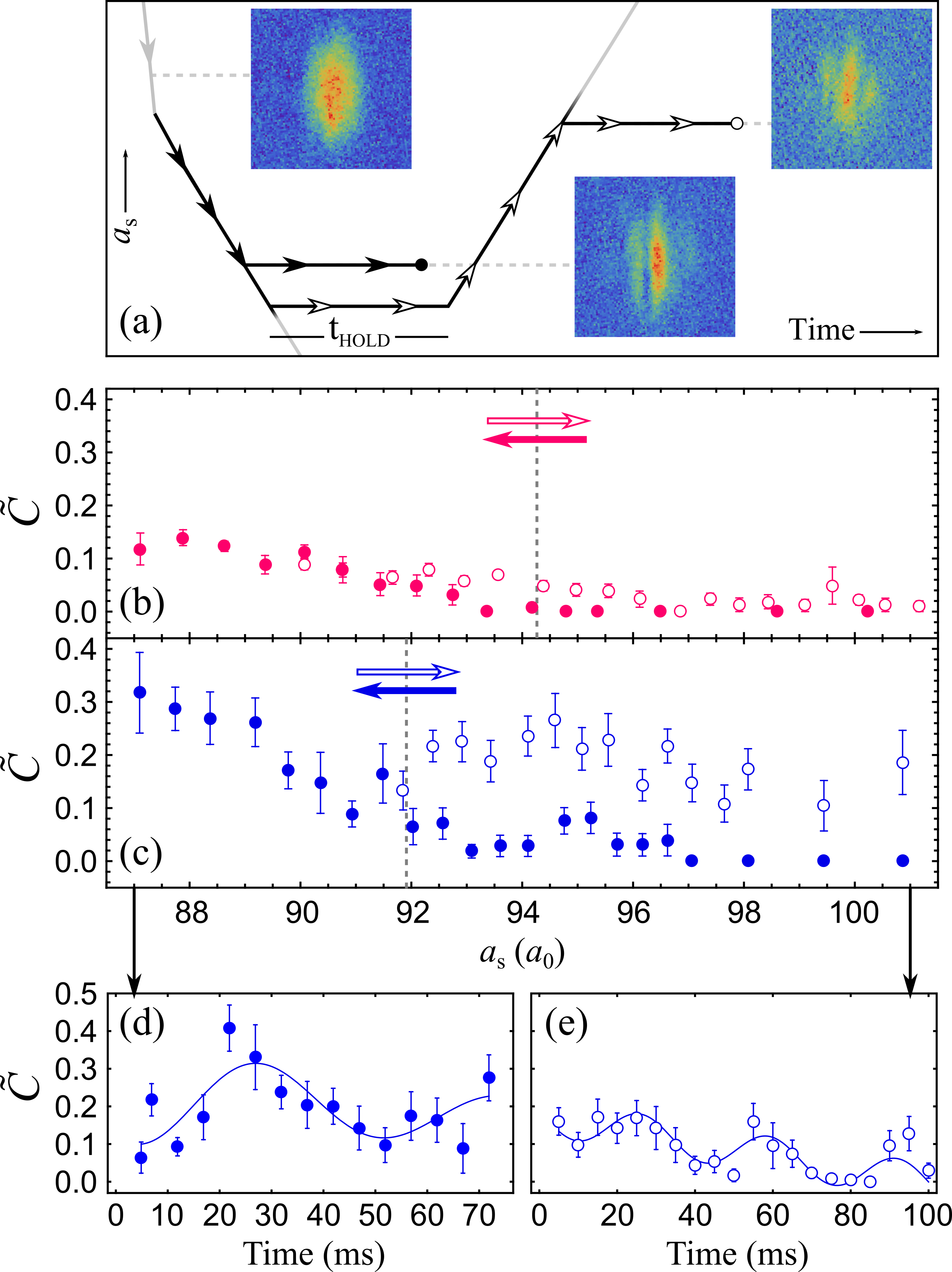}
    \caption{Experimental observation of continuous and discontinuous phase transitions. (a) Trajectories in the $a_s - t$ plane for the in-going (filled symbols) and out-going (empty symbols) ramps, with three samples of the momentum distributions. (b-c) Contrast $\contrast$ vs scattering length during in-going (dots) and out-going (circles) ramps for potentials $V_C$ (b) and $V_D$ (c), respectively. Vertical dashed lines mark the theoretical position of the transitions. (d-e) Time evolution of $\contrast$ for potential $V_D$ at $a_s=87.3$~$a_0$ in the in-going ramp (d) and at $a_s=100.3$~$a_0$ in the out-going ramp (e), respectively. Dots are experimental data, lines are fits with a damped oscillation model. Error bars represent the standard error of the mean of about 10-20 measurements.}
    \label{fig:AndaRianda}
\end{figure}

Unlike the theory presented so far, in the experiment we study the dynamical evolution of the system for varying $a_s$, starting from the superfluid side of the transition. This does not necessarily correspond to following the equilibrium ground state. The dynamical nature of the problem involves concepts not discussed so far, such as adiabaticity, dissipation and, in the case of discontinuous phase transitions, hysteresis.

The experiments start from a standard superfluid, at temperatures well below the critical temperature for condensation. The scattering length is then slowly reduced by means of a magnetic Feshbach resonance. The speed of the magnetic-field ramp was chosen as a compromise between adiabaticity and a not too large impact of the unavoidable three-body losses \cite{Ta19}. The optimal ramp speed, $\dot{a}_s$=0.5~$a_0$/ms, allows to ramp across the transition in potential $V_C$ almost adiabatically.
We note that the three-body loss rate scales as $\rho^2$, so it reaches its maximum value at the density peaks in the supersolid phase. These losses have both a detrimental and a beneficial effect for the study of the phase transition. If on the one hand they decrease $\rho$ over time, on the other hand they introduce a moderate damping of the excitations connected to a local increase in density (see Appendix \ref{sec:exp_n} for details).

We studied the system via absorption imaging after a sudden release from the optical potential, followed by a long free expansion. Just before the release, we rapidly increase the scattering length to a large value ($a_s\simeq$140~$a_0$ in less than 1~ms) in order to reduce the effect of the interactions \cite{Bo19}. The measured distribution, see examples in Fig.~\ref{fig:AndaRianda}(a), can be related to the theoretical momentum distribution, $\rho(k_x, k_y)$, although there are small modifications due to interactions during the expansion. The main observable is the contrast $\contrast$, i.e. the relative height of the peak at the characteristic momentum of the supersolid, as in the theory (see Appendix \ref{sec:exp_cont}).

Fig.~\ref{fig:AndaRianda} presents the main experimental observations when crossing the superfluid-supersolid transition in potentials $V_C$ and $V_D$. In particular, we study the evolution of $\contrast$ for an in-going ramp from the superfluid to the supersolid (filled circles) and for a subsequent out-going ramp from the supersolid to the superfluid (open circles). The experimental trajectories in the $a_s-t$ plane are shown in Fig.~\ref{fig:AndaRianda}(a). The holding time before imaging is 20~ms, which we found to be the characteristic time to form the supersolid.

Phase transitions in potentials $V_C$ and $V_D$  show completely different behaviors, see Figs.~\ref{fig:AndaRianda}(b-c). In potential  $V_C$ (panel b), the in-going ramp shows a smooth increase of $\contrast$, with small shot-to-shot fluctuations.
During the out-going ramp $\contrast$ returns gradually to zero, indicating that the phase transition can be crossed sequentially in the two directions without creating large excitations.

In potential $V_D$ instead (panel c), the in-going ramp shows strong fluctuations of $\contrast$ already before the phase transition, in the region (97-93)~$a_0$, followed by a steep increase of $\contrast$ around 93 ~$a_0$. At the transition we also observe a jump in the atom number due to the increase of density in the supersolid phase, see Appendix \ref{sec:exp_n}.
Remarkably, during the out-going ramp, $\contrast$ remains very large, up to at least 10~$a_0$ in the superfluid regime. While transition $V_C$ can be crossed back and forth almost adiabatically for our ramp speed, crossing transition $V_D$ is manifestly non-adiabatic.
The two behaviors are consistent with continuous and discontinuous phase transitions, respectively.

In potential $V_D$, we observe strong excitations of various collective modes by crossing the transition. In particular, we focus on the oscillation of the order parameter $\contrast$. Figs.~\ref{fig:AndaRianda}(d-e) show the dynamics of $\contrast$ for two selected values of $a_s$ after the in- and out-going ramps. After the in-going ramp (panel d), the oscillation amplitude is smaller than its mean value, so a finite contrast persists throughout the observed time evolution. The oscillation frequency, $\nu$=21(3)~Hz, is consistent with the so-called amplitude mode of the supersolid, already studied in Ref. \cite{Ta19b}. After crossing back the transition (panel e), we still observe a sinusoidal oscillation. However, in this case the oscillation's amplitude is comparable with its mean value, so  $\contrast$ is about zero in its minima. Since $\contrast$ is proportional to the square of the real space contrast $C$, we identify such oscillation with an excited mode of the superfluid with zero mean value of $C$. Such oscillations are also present for potential $V_C$ but with much smaller amplitude, so we can associate them to a non adiabatic crossing of the phase transition from the supersolid to the superfluid. Note that the large values of $\contrast$ for potential $V_D$  in the static measurements of Fig.~\ref{fig:AndaRianda}(c) are due precisely to the presence of such excited mode of the superfluid, since the measurements are performed close to the first maximum of the oscillation.

\begin{figure}
    \centering
    \includegraphics[width=0.48\textwidth]{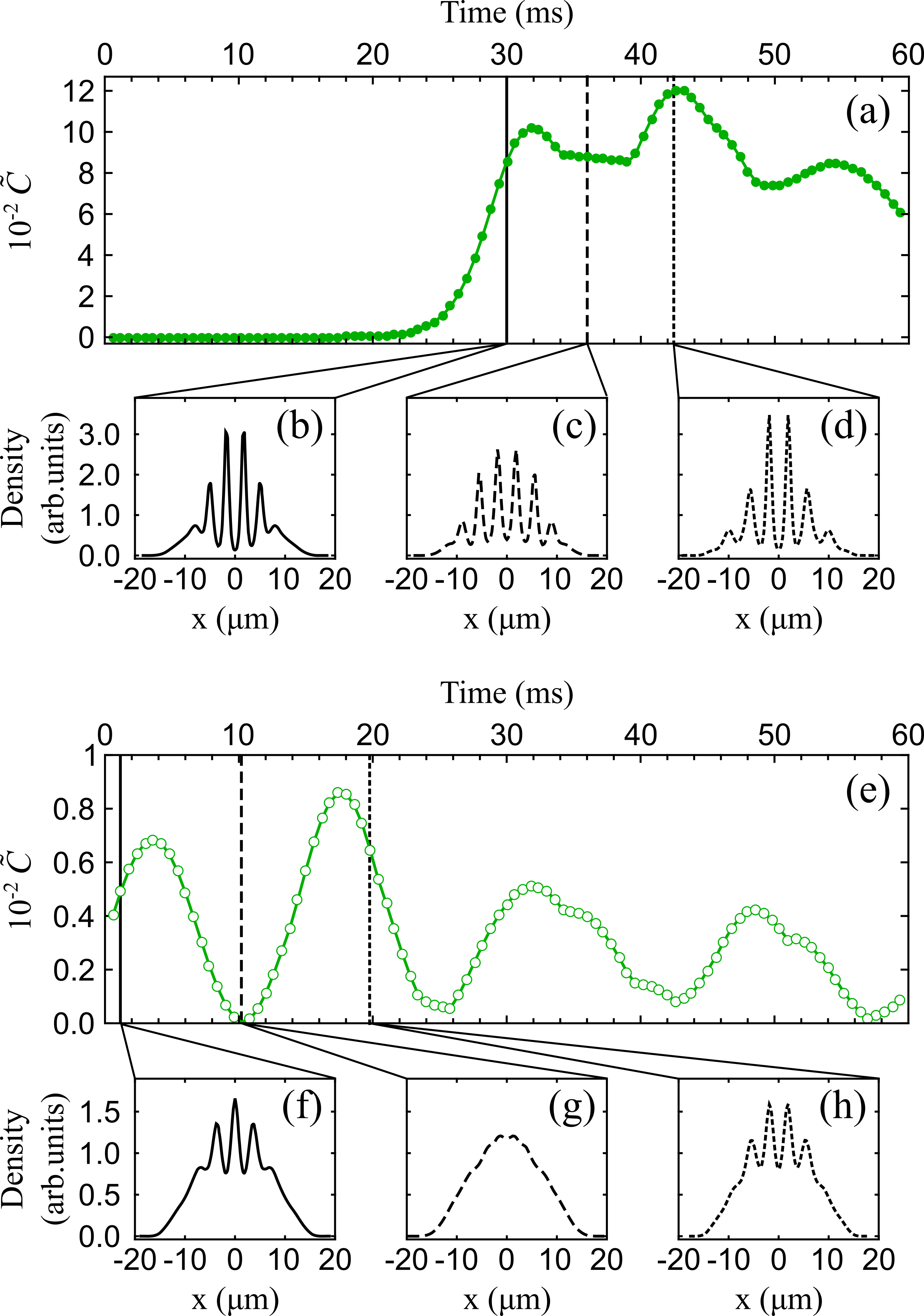}
    \caption{Numerical simulations of the dynamics. Contrast for potential $V_C$ as a function of time, following an in-going ramp from 96~$a_0$ to 93~$a_0$ (a) or an out-going ramp from 93~$a_0$ to 96~$a_0$ (e). Panels (b-d) and (f-h) show examples of the density distribution along $x$ at specific times [vertical lines in (a) and (e), respectively].}
    \label{fig:DynSim}
\end{figure}

To corroborate theoretically these observations, we performed numerical simulations of the dynamics of a simplified zero-temperature system, without quantum fluctuations (besides the LHY energy term), and without losses. Although this system is different from the real one because it lacks dissipation, it allows us to confirm the nature of the observations without introducing phenomenological parameters like temperature or loss rate. Due to the absence of dissipation, the simulations employ scattering-length ramps slower by one order of magnitude than in the experiment, $\dot{a}_s$=0.06~$a_0$/ms, to achieve a quasi-adiabatic crossing of the continuous transition. See Appendix \ref{sec:dynamics} for details.

The simulations confirm the different nature of the observed oscillations. As shown in Fig.~\ref{fig:DynSim}, the in-going ramp excites the amplitude mode of the supersolid, with $\contrast$ featuring small-amplitude oscillations around a relatively large value. The out-going ramp produces instead an oscillation of $\contrast$ with smaller amplitude and with minima at zero, corresponding to an oscillation of the real space $C$ around zero, as shown by the samples of the real-space densities in Fig.~\ref{fig:DynSim}(f-h). Additional simulations show that the latter oscillation can be excited also at fixed scattering length on the superfluid side of the transition, by imposing an initial density modulation similar to that of the supersolid. Furthermore, the oscillation amplitude becomes very small if one removes the dipolar energy term from Eq.~\eqref{eq:TotalEnergy}. This indicates that the oscillations are an excited mode of the superfluid, related to the so-called roton mode \cite{Pe19}, but in a regime of large amplitudes.

\begin{figure}
    \centering
    \includegraphics[width=0.48\textwidth]{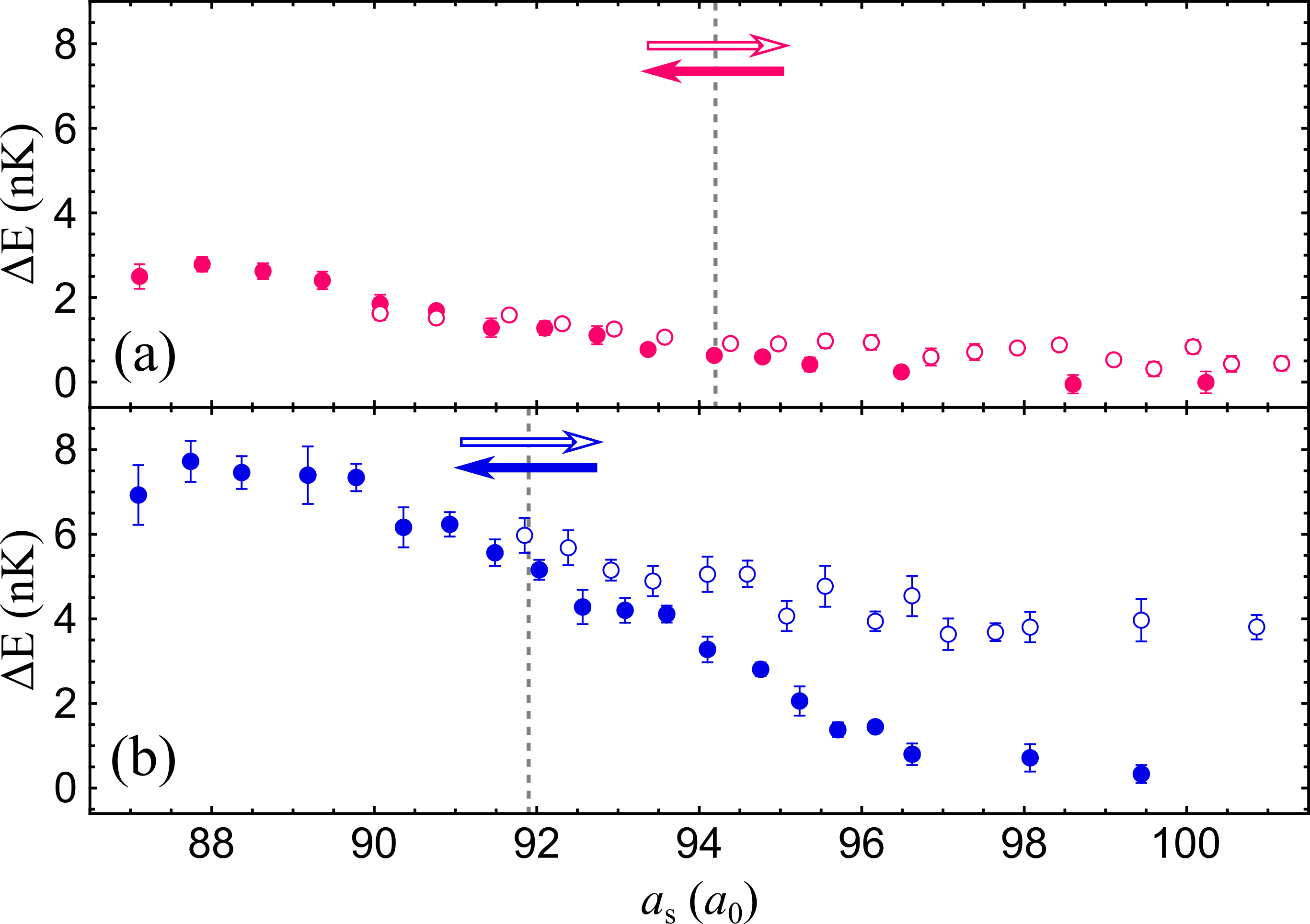}
    \caption{Experimental energy variation across the phase transitions. Variation of the expansion energy for potential $V_C$ (a) and potential $V_D$ (b), during the in-going (filled circles) and out-going  (empty circles) ramps. Error bars are the standard error of the mean of 10-20 measurements.}
    \label{fig:KinEn}
\end{figure}

In  the experiment, crossing the phase transition produces other types of excitations besides the specific amplitude mode analyzed so far, with amplitudes systematically larger in potential $ V_D $ than in potential $ V_C $. In particular we observe the so-called lattice mode of the supersolid \cite{Ta19}, as well as additional longitudinal and transverse modes. In order to quantify the different degree of excitation for the two potentials, we measured the total energy in the momentum distribution after the free expansion, $E_{exp}=\hbar^2\langle k_x^2+k_y^2 \rangle/(2m)$. This quantity overestimates the total energy of the system immediately after the removal of the potential, due to the subsequent increase of $a_s$ during the first phase of the expansion. Fig.~\ref{fig:KinEn} shows the variation $\Delta E$ of the expansion energy with respect to the superfluid with the largest $a_s$. The stronger excitation for potential $V_D$ for both in- and out-going ramps is apparent. A quantitative comparison of the two in-going ramps is difficult, since the supersolid in potential $V_D$ has generally a larger contrast, which implies larger $E_{exp}$. A comparison can instead be made for the out-going ramps, since both end in the superfluid regime. Here, potential $V_D$ shows an excess energy $\Delta E\simeq4$~nK, much larger than in potential $V_C$, $\Delta E\simeq0.5$~nK. This is a further evidence of the different character of the phase transitions in the two potentials.  The numerical simulations show a similar difference in excess energy when crossing the two phase transitions, $\Delta E\simeq0.5$~nK for potential $V_C$ and $\Delta E\simeq2.5$~nK for potential $V_D$, see Appendix~\ref{sec:dynamics} for details.

\begin{figure*}
    \centering
    \includegraphics[width=0.95\textwidth]{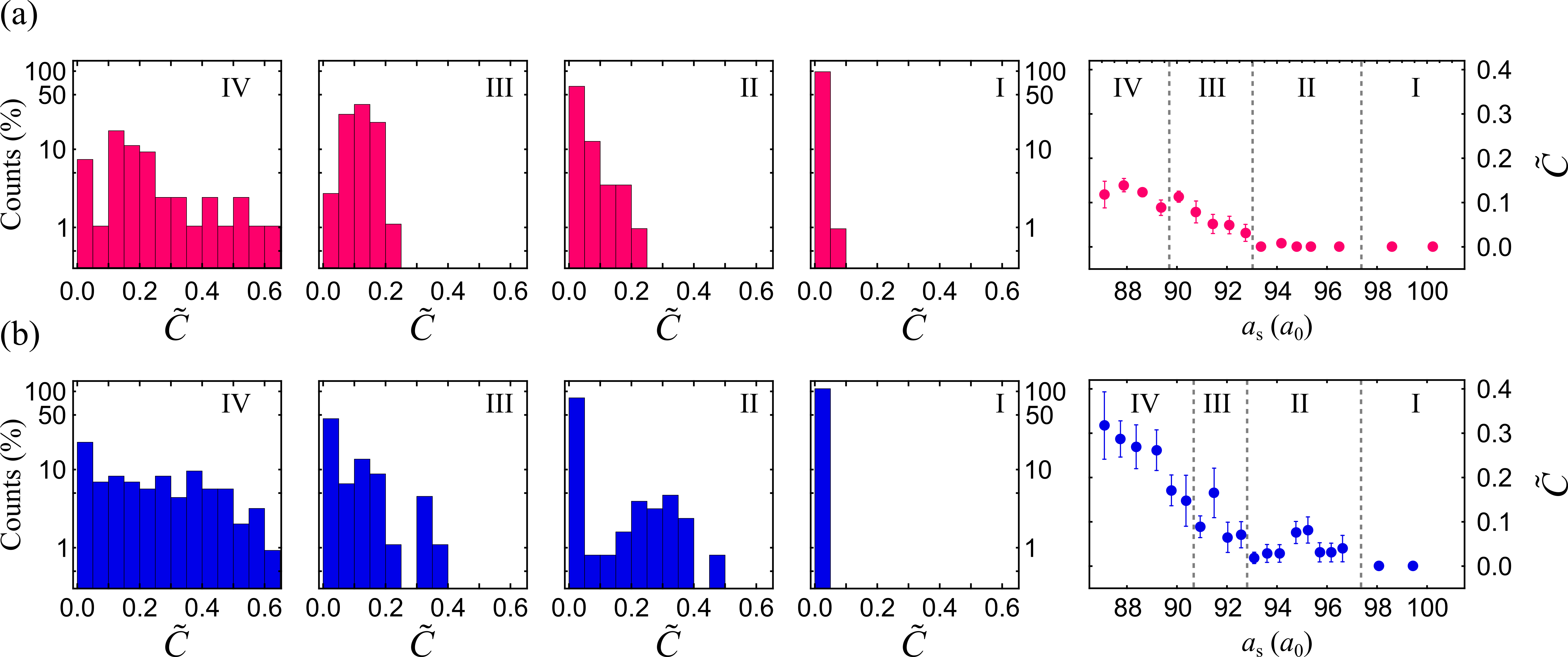}
    \caption{Character of the phase transition from fluctuation spectra. Histograms of the contrast fluctuations for potential $V_C$ (a) and for potential $V_D$ (b), respectively. Bins correspond to different regions of scattering length, accordingly to the labels (I-IV). Each bin contains 100-150 samples. Single- and double-peak structures in the central bins for potential $V_C$ and $V_D$, respectively, demonstrate the different shape of the free energy. }
    \label{fig:Fluc}
\end{figure*}

A relevant question is whether the discontinuous transition in potential $V_D$ shows hysteresis similarly to first-order phase transitions, i.e. different locations of the phase transition depending on the direction in which the transition is crossed. However, the concept of hysteresis (in the thermodynamic sense) applies only to systems that are in equilibrium. Our system is out of equilibrium, as it cannot entirely dissipate the energy acquired when crossing the transition from the superfluid to the supersolid. Therefore, we do not expect to observe hysteresis. In the dynamical measurements as in Fig.~\ref{fig:AndaRianda}(d-e) we indeed do not observe any hysteresis, within our current resolution of 2~$a_0$. The simulations predict an hysteresis of just about 0.5~$a_0$ in the hypothetical equilibrium case, see Appendix~\ref{sec:dynamics}.

\section{\label{sec:fluctuations}Experimental fluctuation spectrum}

A confirmation of the different character of the two phase transitions comes from an analysis of the fluctuations of the contrast. Fluctuations are generally enhanced in the vicinity of a phase transition, including the superfluid-supersolid phase transition \cite{He21}. We already noted the presence of shot-to-shot fluctuations of $\contrast$ for the data in Fig.~\ref{fig:AndaRianda}(a-b). These fluctuations can have quantum, thermal or technical origin, the latter being presumably dominated by the shot-to-shot fluctuations in the atom number, which in turn determine fluctuations in the critical scattering length. One expects the fluctuations to have different distributions for continuous and discontinuous phase transitions, due to the different shape of the Landau free energy.

Fig.~\ref{fig:Fluc} presents the distributions of $\contrast$ measured for the two transitions, binned in four intervals of scattering length: well before, just before, just after and well after the transition. For potential $V_C$, the distributions show the expected behavior of a continuous phase transition, see Fig.~\ref{fig:Fluc}(a). For large $a_s$ (region I), the system is in the superfluid phase in each experimental run. For smaller $a_s$, when the modulation appears (region II), a portion of the samples comes apart from the peak at zero contrast, populating a region of small contrasts nearby. Deeper into the supersolid region (region III), the peak at zero contrast is depleted and the system occupies a well-defined set of contrasts. Decreasing further $a_s$ towards the droplet crystal regime (region IV), the contrast becomes larger on average but also noisier, probably because of the enhancement of three-body losses which modify from shot to shot the density distribution, depending on the details of the dynamical formation process.

For potential $V_D$, see Fig.~\ref{fig:Fluc}(b), the situation is quite different. Just before the transition (region II), the histogram separates into two peaks: a main one at zero contrast and a smaller one at finite contrast. We interpret the latter peak as due to fluctuations into the second minimum of the Landau free energy, across the barrier. The double-peak structure persists in the next bin (panel III), while moving further towards the droplet crystal one recovers the same broad distribution of the continuous case.

A double-peak structure of the order parameter is a signature of first-order phase transitions, see e.g. \cite{Bo11}, and is associated to the coexistence of the two competing phases. In our case, the comparison of the double peak in Fig.~\ref{fig:Fluc}(b) with the single peak in Fig.~\ref{fig:Fluc}(a) gives evidence of the different energy landscapes for the two experimental configurations. This provides a conclusive evidence of the different character of the two phase transitions, continuous for potential $V_C$ and discontinuous for $V_D$.

\section{\label{sec:conclusions}Discussion and conclusions}

In summary, we assessed the character of the superfluid-supersolid quantum phase transition, finding a crossover from 1D-like configurations with continuous phase transitions into 2D-like configurations with discontinuous phase transitions. The crossover is due to the very nature of the supersolid. Although our supersolids have only a single row of principal density maxima, they keep a 2D structure thanks to the presence of the background density associated to their superfluid nature. Along the crossover, the 2D structure is gradually suppressed by an increasing transverse confinement or an increasing atom number. Our analysis establishes also a link between previous numerical simulations for infinite systems, showing that the discontinuous transition seen in quasi-1D at low density \cite{Bl20} has the same nature of the discontinuous transition in 2D \cite{Zh19}.

The continuous-discontinuous crossover is controlled experimentally by changing the transverse confinement. We  achieve various evidences of the different character of the transition on the two sides of the crossover, in particular from the striking difference in the adiabaticity with which the phase transition can be crossed, besides the different shape of the fluctuation spectrum. The continuous transition can be crossed almost adiabatically despite the limited time scales of the experiment, with a residual energy increase of just about 0.5 nK. The discontinuous transition is quite sharp despite the very small length of the supersolid lattice, which has only 2-3 sites. That supports the expectation that broadening effects are determined by the number of particles ($N\simeq10^4$ here) and not by the number of sites \cite{Im80}.

Our findings have implications for future studies of the supersolid phase of matter. Crossing continuous quantum phase transitions we achieve supersolids that are almost free of excitations.
This is an important prerequisite to study a key property of supersolids, the reduced superfluidity due to the crystal-like structure \cite{Le70}. Our results can be extended directly to annular configurations, which are the ideal setup to study superfluidity \cite{Le70,Ro19,Sa20} and would also allow to eliminate the longitudinal harmonic potential without the edge effects introduced by box-like potentials \cite{Ro21}. A sufficiently tight radial confinement will allow to achieve continuous phase transitions, in analogy with the 1D-like configurations we explored in this work. The control of continuous quantum phase transitions opens up also the possibility to study the generation of entanglement in the supersolid phase. Indeed, the increase of correlations in a quantum transition can lead to the creation of many-body entanglement \cite{Xi17}. Crossing adiabatically a continuous transition would limit the impact of noise and decoherence which usually leads to degradation of fragile entangled states. Understanding quantum correlations in the supersolid would be an important step both for developing a full quantum description of this new phase of matter and for the possible exploitation of metrologically-useful entangled states \cite{Pe18}.

%------------
%Acknowledgments
%------------
\begin{acknowledgments}
We acknowledge discussions with I. L. Egusquiza and A. Smerzi.
This work was supported by the EC-H2020 research and innovation program, through the project 641122-QUIC, by the QuantERA grant MAQS, by CNR-INO, by Grant PGC2018-101355-B-I00 funded by MCIN/AEI/10.13039/501100011033 and by “ERDF A way of making Europe”, and by the Basque Government through Grant No. IT986-16.
\end{acknowledgments}

\appendix

\section{\label{app:landau_model}Landau model for the  dipolar supersolid}
In this section we discuss in more detail the superfluid-supersolid phase transition in the framework of the Landau model.
A quantitative comparison is not possible because the Landau model assumes a fixed lattice geometry, and therefore cannot capture the effects arising from the competition between supersolid compressibility and trapping potential, which determine the phase diagram in Fig~(\ref{fig:PhaseDiagram}). Neverthless, it is useful to get a qualitative understanding of the dimensional crossover.

First, we consider the case of a 1D density modulation of the form $\rho (x)=\rho_0 \big[1+ C \cos ( k x) \big] $, which provides a good characterization of the one-row configuration. Inserting this ansatz into the energy functional Eq.~(\ref{eq:TotalEnergy}) and expanding in powers of $C$ leads to a Landau expansion of the form of Eq.~(\ref{eq:LandauExpansion}), where the coefficients depend on the wavevector $k$, the mean density $\rho_0$, and the scattering length $a_s$. The odd terms are exactly vanishing, as a consequence of the symmetry in the sign of the contrast discussed in the main text. Both contact and dipolar energies contribute only to the quadratic coefficient $b_{1D}$, since they are quadratic in the density. The coefficient $b_{1D}$ decreases as the scattering length is decreased, and becomes negative when the attractive part of the dipolar interaction dominates, favouring the formation of the supersolid state. The other coefficients are determined only by the kinetic and LHY energies, which are more complicate functions of the density. The fourth- and the sixth-order coefficients $d_{1D}$ and $e_{1D}$ read
\begin{equation}
    \begin{split}
         d_{1D} & = \frac{\hbar^2 \pi}{32 m}k\rho_0-\frac{15 \pi g_{lhy}}{512 k}\rho_0^{5/2}\,, \\
         e_{1D} & = \frac{\hbar^2 \pi}{64 m}k\rho_0-\frac{25 \pi g_{lhy}}{8192 k}\rho_0^{5/2}\,,
    \end{split}
\end{equation}
where $g_{lhy}$ is defined after Eq.~(\ref{eq:GPenergy}).
In the experimental density regimes, the contribution of the kinetic energy (first term on the right-hand sides) dominates over that of the LHY (second term). The coefficient $d_{1D}$ is then positive and the next-order term does not change the scenario of a continuous phase transition. When instead the density is large enough, quantum fluctuations overcome the kinetic energy reversing the sign of $d_{1D}$. To ensure stability of the system, one has to consider the term $e_{1D}~C^6$. The resulting Landau energy $\Delta E = b_{1D}~C^2 + d_{1D}~C^4 + e_{1D}~C^6$ predicts a discontinuous phase transition with the combination $b_{1D}>0,~d_{1D}<0,~e_{1D}>0$. The shape of the energy is depicted in Fig.~(\ref{fig:Fig1})(b). This mechanism justifies the discontinuous phase transitions at high densities and high trap frequencies depicted in gray in Fig.~(\ref{fig:PhaseDiagram}).

The previous simple model can be extended to include the effects of the trap. We modify the density ansatz into $\rho (x) =  \rho_0 g(x) \big[1+ C \cos ( k x) \big]$, where $g(x)$ is a gaussian envelope of width $\sigma$ that contains the proper normalization constant. As a result of the presence of the envelope, the symmetry in the sign of the contrast is broken. The two density distributions with $C$ and $-C$ are indeed slightly different, since they have the main maximum and the central minimum at the center of the trap, respectively. Consequently, odd terms appear in the Landau energy. We have checked that odd terms are below 10\% of the even terms in the relevant regime in which more than one cluster is present, and approach zero when many clusters are present, i.e. when $k\sigma$ is large.  The presence of the trap introduces also linear terms in the contrast. In particular, the trap energy is proportional to the density, so it contributes only to the linear term. The interpretation of this fact is clear: a state with $C>0$ is favorable in trap energy compared to a superfluid with $C=0$, since it increases the density in the center of the trap and forms lateral minima where the trap potential is higher. The presence of a trap-induced linear term in the Landau energy can explain the mixing of continuous and discontinuous phase transitions observed in the simulations. Indeed, in configurations where the transition is discontinuous, the presence of a linear term can induce a continuous transition towards states of small contrast, followed by a jump towards the second minimum at higher contrast.

In the 2D infinite case, the full calculation of the Landau energy with a sinusoidal ansatz of the form Eq.~(\ref{eq:Density}) was reported in Ref. \cite{Zh19}.
We checked to what extent this model applies to our case with an harmonic potential in the $y$ direction, modifying the ansatz into $\rho (x,y) = \rho_{2D} (x,y) g(y)$, where $\rho_{2D}$ is given by Eq.~(\ref{eq:Density}) and $g(y)$ is a gaussian envelope in the transverse direction $y$ of width $\sigma$. The expansion coefficients are equal to the thermodynamic limit in 2D until $\sigma k/ (2 \pi)$ approaches 0.4, a regime in which the first lateral rows of clusters are already suppressed to approximately 10\%. This is not dissimilar from the results of the numerical simulations in Fig.~(\ref{fig:DimensionalOrigin}), which show the crossover to continuous transitions when the lateral rows are suppressed to a similar level. Such agreement confirms the origin of the dimensional crossover discussed in the main text.

\section{\label{sec:methods_num}Numerical simulations of the phase transition in equilibrium}

\subsection{Numerical methods}

In this section we present the methodology followed for the numerical simulations of the equilibrium phase diagram, and we discuss the additional analysis employed to characterize the phase diagram.
The system is described in terms of a generalized Gross-Pitaevskii (GP) theory including the beyond mean field Lee-Huang-Yang correction. The energy functional is $E = E_{mf} + E_{dd} + E_{LHY}$ with

\begin{align}
E_{mf} &=
\int \left[\frac{\hbar^2}{2m}|\nabla \psi(\bm{r})|^2  + V_{ho}(\bm{r})\rho(\bm{r})+\frac{g}{2} \rho^2(\bm{r})
\right]d\bm{r}\,,
\nonumber\\
E_{dd} &=\frac{C_{dd}}{2}\iint \rho(\bm{r})V_{dd}(\bm{r}-\bm{r}')\rho(\bm{r}') d\bm{r}d\bm{r}'\,,
\nonumber\\
E_{LHY} &=g_{lhy}\int \rho^{5/2}(\bm{r})d\bm{r}\,,
\end{align}\label{eq:GPenergy}

where  $\rho(\bm{r})=|\psi(\bm{r})|^2$ represents the condensate density, $V_{ho}(\bm{r})=(m/2)\sum_{\alpha=x,y,z}\omega_{\alpha}^{2}r_{\alpha}^{2}$ the harmonic trapping, $g=4\pi\hbar^2 a_{s}/m$ is the contact interaction strength,
$V_{dd}(\bm{r})= (1-3\cos^{2}\theta)/(4\pi r^{3})$ the (bare) dipole-dipole potential,
$C_{dd}\equiv\mu_{0}\mu^2$ its strength, $\mu$ the modulus of the dipole moment $\bm{\mu}$, $\bm{r}$ the distance between the dipoles, and $\theta$ the angle between the vector $\bm{r}$ and the dipole axis, $\cos\theta=\bm{\mu}\cdot\bm{r}/(\mu r)$ \cite{Ronen:2006}. The orientation of the magnetic dipoles is along the $z$ direction (the direction of the magnetic field $\bm{B}$). The LHY coefficient is $g_{lhy}=\frac{256\sqrt{\pi}}{15}\frac{\hbar^{2}a_s^{5/2}}{m}\left(1 + \frac{3}{2}\epsilon_{dd}^{2}\right)$, with $\epsilon_{dd}=\mu_0 \mu^2 N/(3g)$\cite{Wachtler:2016}.

The ground state of the system is obtained by minimizing the energy functional $E[\psi]$ by means of a conjugate algorithm, see, e.g., Refs.\,\cite{press2007,Modugno:2003,Ronen:2006}. In the numerical code the double integral appearing in Eq.\,(\ref{eq:GPenergy}) is mapped into  Fourier space where it can be conveniently computed by means of fast Fourier transform (FFT) algorithms, after regularization.
The LHY correction in Eq.\,(\ref{eq:GPenergy}) is obtained from the expression for homogeneous 3D dipolar condensates under the local-density approximation \cite{Wachtler:2016,Schmitt:2016}.

To obtain the contrast in Fourier space $\contrast$ from the ground state density distributions, we first compute $|\mathcal{F}[\sqrt{\rho\left(x\right)}]|^2$, where $\rho \left(x\right)$ is the column density integrated along the transverse trap direction $y$. The order parameter $\contrast$ is given by the height of the first lateral peak relative to the central one. A simple relation between real space $C$ and $\contrast$ can be derived in the 1D infinite case discussed in Appendix~\ref{app:landau_model}. In the limit $C\ll1$, we find $\contrast=C^2/16$.

\subsection{\label{sec:datanum}Additional data of numerical simulations}

\begin{figure}[b]
    \centering
    \includegraphics[width=0.48\textwidth]{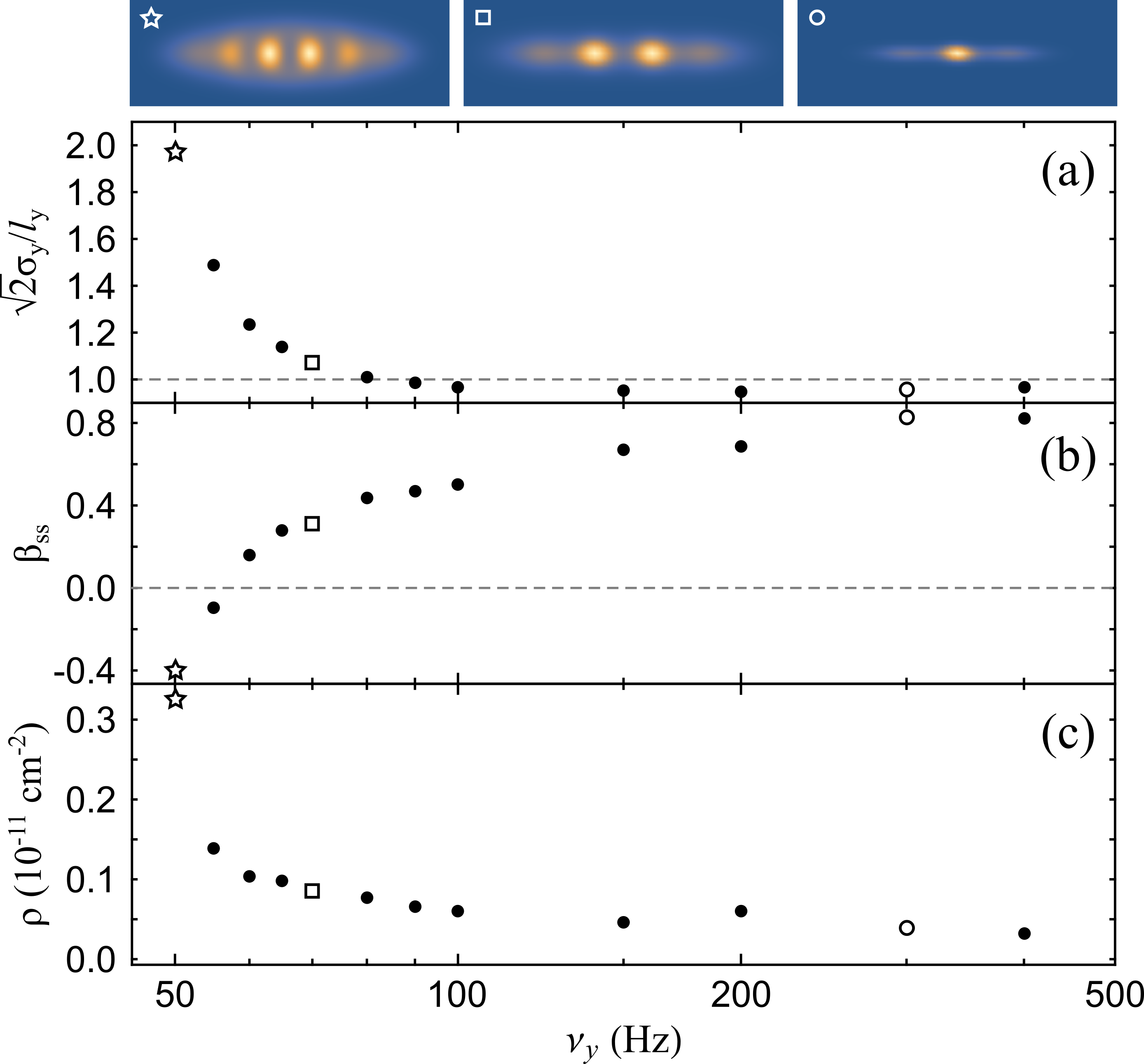}
    \caption{Characterization of the supersolid structure for the transitions at the center of the dimensional crossover. (a) Transverse width $\sigma_y$, (b) deformation parameter $\beta_{ss}$ of the central supersolid clusters and (c) peak density vs transverse trap frequency. Insets show examples of the supersolid density distributions in the $x- y$ plane.}
    \label{fig:DataCrossover}
\end{figure}

Fig.~\ref{fig:DataCrossover} shows data supporting the discussion in Section~\ref{sec:numerical} about the compressibility of the supersolid and the onset of the LHY-dominated regime, for the transitions at the center of the crossover. Panel (a) shows the transverse width $\sigma_y$ normalized to that of a non-interacting system, $\ell_y/\sqrt{2}$. For $\nu_y > 70$~Hz the values are very close to $1$, in agreement with the expectation that, at the transition, attractive dipolar and repulsive contact interactions tend to cancel out. Since $\sigma_y$ is slightly less than $\ell_y/\sqrt{2}$, the dipolar interaction is slightly larger than the contact interaction. For smaller frequencies, instead, one notes a very rapid increase of the peak density [panel (c)]. This brings the system into a regime in which the LHY energy has an important role in limiting further density increases, due to its $\rho^{5/2}$ dependence, by increasing the transverse size of the clusters. So, $\sigma_y$ becomes larger than the non-interacting width. In panel (b), we observe the same features in the deformation parameter $\beta_{ss}=\langle x^2-y^2\rangle/\langle x^2+y^2\rangle$ of the central supersolid clusters. In an unconfined 2D system, the natural shape of the clusters in the $x- y$ plane is circular, so that $\beta_{ss}=0$. In the trapped system studied in the main text, this happens only in a limited range of transverse trapping frequencies around $\nu_y$=60~Hz. For larger frequencies, the clusters get squeezed by the transverse trap, so one has $\beta_{ss}>0$. For smaller frequencies, in the LHY regime, the clusters become elongated in the transverse direction and $\beta_{ss}<0$. This is the ``stripe'' regime found also in other numerical studies \cite{Zh21,He21b}.

\begin{figure}
    \centering
    \includegraphics[width=0.48\textwidth]{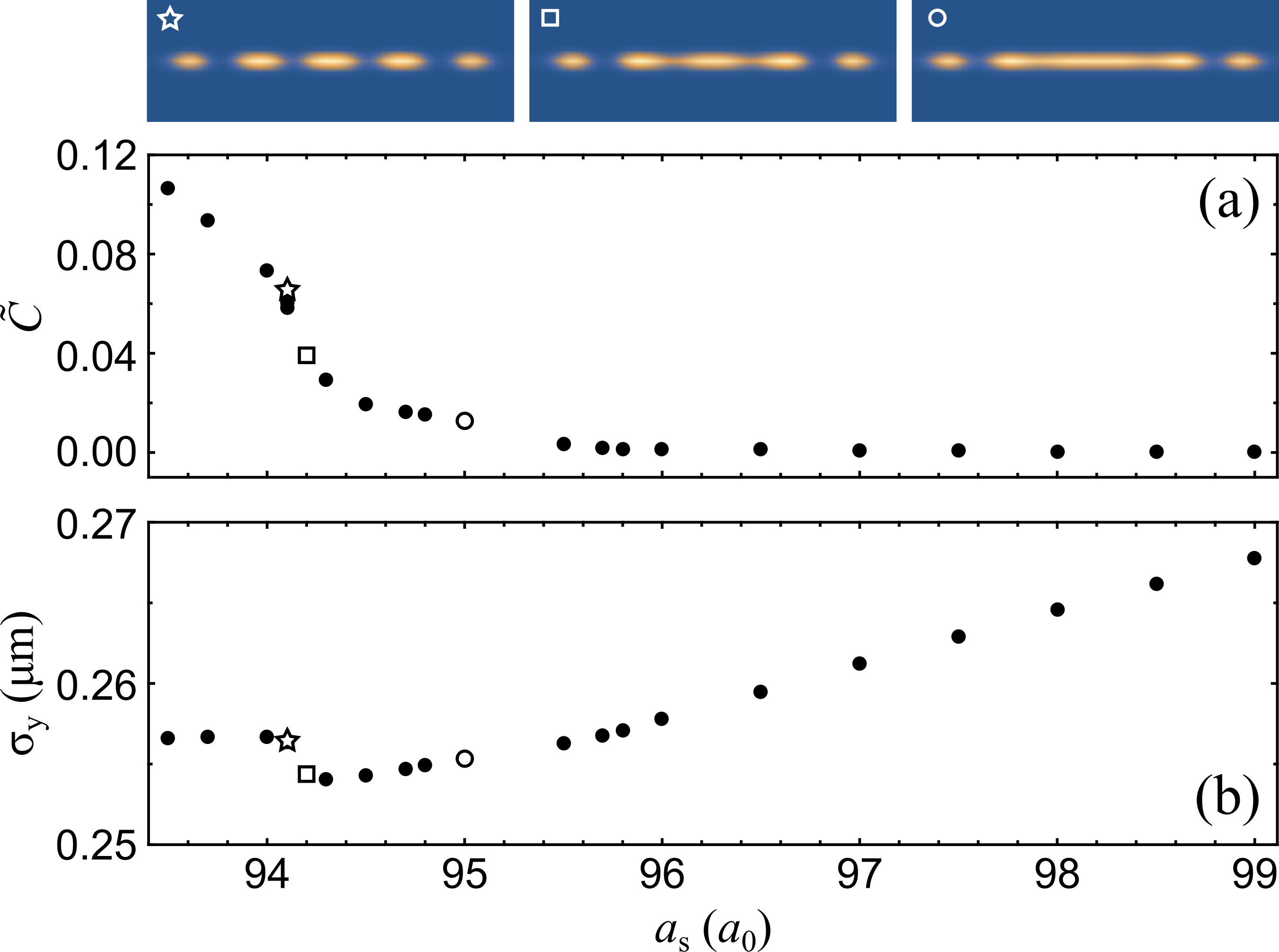}
    \caption{Discontinuous transition in the LHY-dominated regime, ($\nu_y$=400 Hz, $N$=7$\times 10^4$). (a) Contrast $\contrast$ as a function of the scattering length. Note the initially continuous transition around 95.5~$a_0$, followed by a discontinuous one around 94.2~$a_0$. The insets show that the modulation initially forms on the sides of the distribution. (b) Behavior of $\sigma_y$ across the transition, demonstrating the inverse magnetostriction provided by the LHY term.}
    \label{fig:DataCrossoverb}
\end{figure}

A different effect of the LHY energy term arises for large transverse confinements and large $N$, leading to a second crossover from continuous to discontinuous transitions, as shown in Fig.~\ref{fig:PhaseDiagram} (a). Fig.~\ref{fig:DataCrossoverb} shows an example of the discontinuous transition in such regime. The characteristic features are: 1) the supersolid initially forms continuously in the regions with intermediate density, where the LHY term is not too large, and not at the center as usual (a similar effect was observed numerically in Ref. \cite{Ch19}); 2) in a second stage, the supersolid forms also in the center, via a discontinuous transition since the LHY term there is large; 3) the discontinuous transition is accompanied by the opposite of the standard magnetostriction effect, i.e. the transverse size increases when the supersolid forms, to reduce the density increase; 4) the discontinuous transition is weak, because it regards only a small fraction of the system. All these effects are a consequence of the peculiar dependence of the LHY energy term on the density, $\rho^{5/2}$, which results in a stronger variation of the LHY term along the system than the contact and dipole-dipole terms.

\section{\label{sec:methods_exp}Experimental methods and analysis}

\subsection{\label{sec:exp_seq}Experimental sequence}
The experiments start with a Bose-Einstein condensate of about 8$\times 10^4$ $^{162}$Dy atoms in a crossed optical trap at a magnetic field $B\simeq5.5$~G, corresponding to a scattering length of about 140~$a_0$. The magnetic field is then slowly changed towards the critical values for the phase transition into the supersolid, close to the set of Feshbach resonances around 5.3~G \cite{Ta19}, with a resolution of 1~mG. The magnetic field is calibrated via radio-frequency spectroscopy before and after each experimental run, lasting typically 3-4 hours. The magnetic field stability is about 0.5~mG, corresponding to a stability in scattering length of about 0.3~$a_0$. For the conversion from magnetic field to scattering length, we adopt the model presented in \cite{Bo19b}. The relatively large uncertainty in the background scattering length results in a global systematic uncertainty of several $a_0$ in the experimental scattering length at which the transitions take place. In all experimental figures we applied a shift to the experimental scattering length to get a matching between the experimental and theoretical transitions in potential $V_D$. We identify the experimental transition point from the jump in the atom number, see Fig.~\ref{fig:expN}. The resulting shift amounts to $6.7~a_0$.

At the end of each experimental sequence, we suddenly switch off the optical potential and we let the system expand for 90~ms in the presence of a magnetic-field gradient that compensates gravity. About 200~$\mu$s before the release of the atoms we increase the contact interaction strength by setting $a_s\simeq 140~a_0$, thus minimizing the effects of the dipolar interaction on the expansion. We finally measure the atomic density, which we interpret as the momentum distribution $\rho(k_x,k_y)$, by absorption imaging on the strong optical transition at 421~nm.

\subsection{\label{sec:exp_cont}Analysis of the contrast}
\begin{figure}
    \centering
    \includegraphics[width=0.48\textwidth]{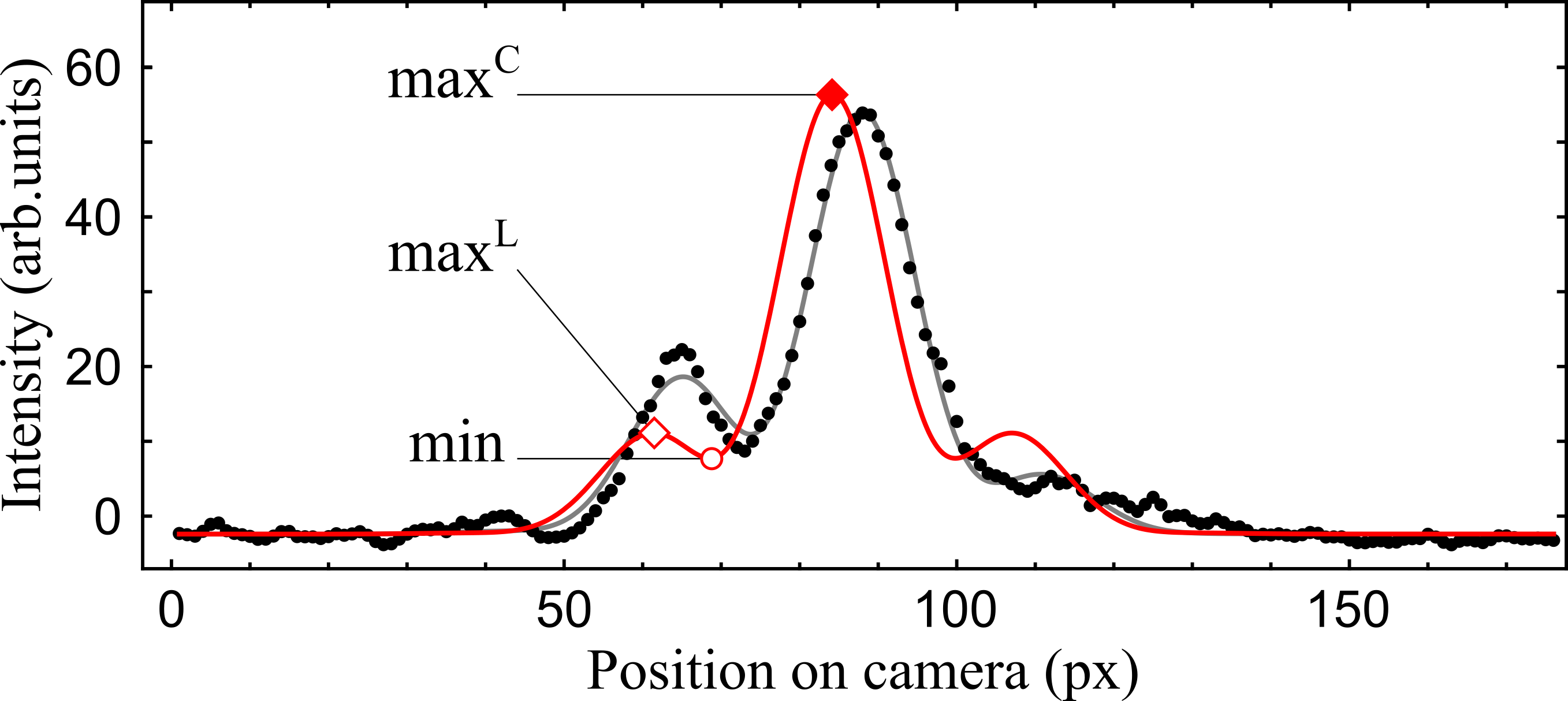}
    \caption{Example of experimental 1D momentum distribution $n(k_x)$ obtained after a time-of-flight experiment in the supersolid regime. Black dots are experimental data, while the gray line is the fit using the double-slit function of Eq.~(\ref{eq:Fit}). The red line is the same fitted function after fixing the phase to $\pi/2$. The highlighted points are the central maximum $\mathrm{max}^C$, the lateral maximum $\mathrm{max}^L$ and the intermediate minimum $\mathrm{min}$ used in the definition of the observables $\contrast$ and $\contrast_2$ (see text).}
    \label{fig:ExpContrast}
\end{figure}

The presence of the supersolid density modulation is revealed by the characteristic side peaks in the momentum distribution, as depicted in the insets of Fig.~\ref{fig:AndaRianda}(a). To extract the contrast $\contrast$, both as a function of the scattering length and of time, we analyze each image following these steps. First, we rotate the momentum distribution in the plane to align the interference peaks along the $k_x$ direction. Second, we integrate over $k_y$ to get the 1D momentum distribution $\rho(k_x)$. Third, we fit $\rho(k_x)$ with a double-slit model
\begin{equation}
\label{eq:Fit}
     \rho(k_x) = A_0~ \esp{-\frac{(k_x-k_0)^2}{2 \sigma^2}} \left\{1+A_1\sin\left[{\pi(k_x-k_0)}/{k_r} + \phi\right] \right\}
\end{equation}
where $A_0, k_0$ and $\sigma$ are, respectively, the amplitude, center and width of the envelope, while $A_1, k_r$ and $\phi$ are the amplitude, period (in the momentum space) and phase of the modulation. The typical experimental image has a phase of $\pi/2$, corresponding to a central peak and two lateral, symmetric peaks. To get a contrast $\contrast$ unbiased from fluctuations of the phase around $\pi/2$, we rephase the fitted function imposing $\phi=\pi/2$ and we define $\contrast = \mathrm{max}^L / \mathrm{max}^C$, where $\mathrm{max}^C$ and $\mathrm{max}^L$ are the values of the central and first lateral maxima, respectively (see Fig.~\ref{fig:ExpContrast}). Since the lateral maximum in the superfluid phase does not exist, we should get $\contrast=0$ in this regime. However, trying to fit a superfluid profile with the double-slit function Eq.~(\ref{eq:Fit}), the fit can force the presence of small maxima in the slope of the superfluid and give fictitious non-zero values of $\contrast$. To overcome this problem, we use a different observable defined as $ \contrast_2 = (\mathrm{max}^L-\mathrm{min})/(\mathrm{max}^L + \mathrm{min})$, where $ \mathrm{min}$ is the first lateral minimum, and we set $ \contrast = 0$ if $ \contrast_2 = 0$. The reason is that the observable $ \contrast_2$ is automatically zero in the superfluid phase, since eventual fictitious lateral peaks practically coincide with their corresponding minimum. However, we don't take $ \contrast_2$ as a measurement of the contrast since it is very sensitive to the atom number, which is lower and lower entering the supersolid state due to three body-losses. Typically, $\contrast_2$ has a maximum lowering the scattering length and then start decreasing. Thus, we use $\contrast_2$ only to distinguish between superfluid and supersolid images.

When we plot the contrast as a function of time, the fitting function is a damped sinusoid of the form
\begin{equation}
    \contrast(t) = A \sin{\big( \sqrt{(2\pi\nu)^2-\tau^2} ~t+\phi \big)}e^{-t/\tau}+\text{off}_1 t +\text{off}_2
\end{equation}
where $\nu$ is the frequency of the oscillation and $\tau$ the damping time. We fit with the same function also other observables which allow a direct comparison between the BEC and the supersolid, for example the longitudinal width $\sigma_x$ which features the breathing-mode oscillation \cite{Ta19b}. From measurements performed at different values of the scattering length, see e.g. Fig.~\ref{fig:AndaRianda}(d-e), we find that the damping time decreases of about an order of magnitude going from the BEC to the supersolid. For example, in the discontinuous potential $V_D$ we get $\tau=100\pm 41$~ms at $94.3~a_0$ and $\tau=15\pm 5$~ms at $87.4~a_0$.

\subsection{\label{sec:exp_n}Analysis of the atom number}
\begin{figure}
    \centering
    \includegraphics[width=0.48\textwidth]{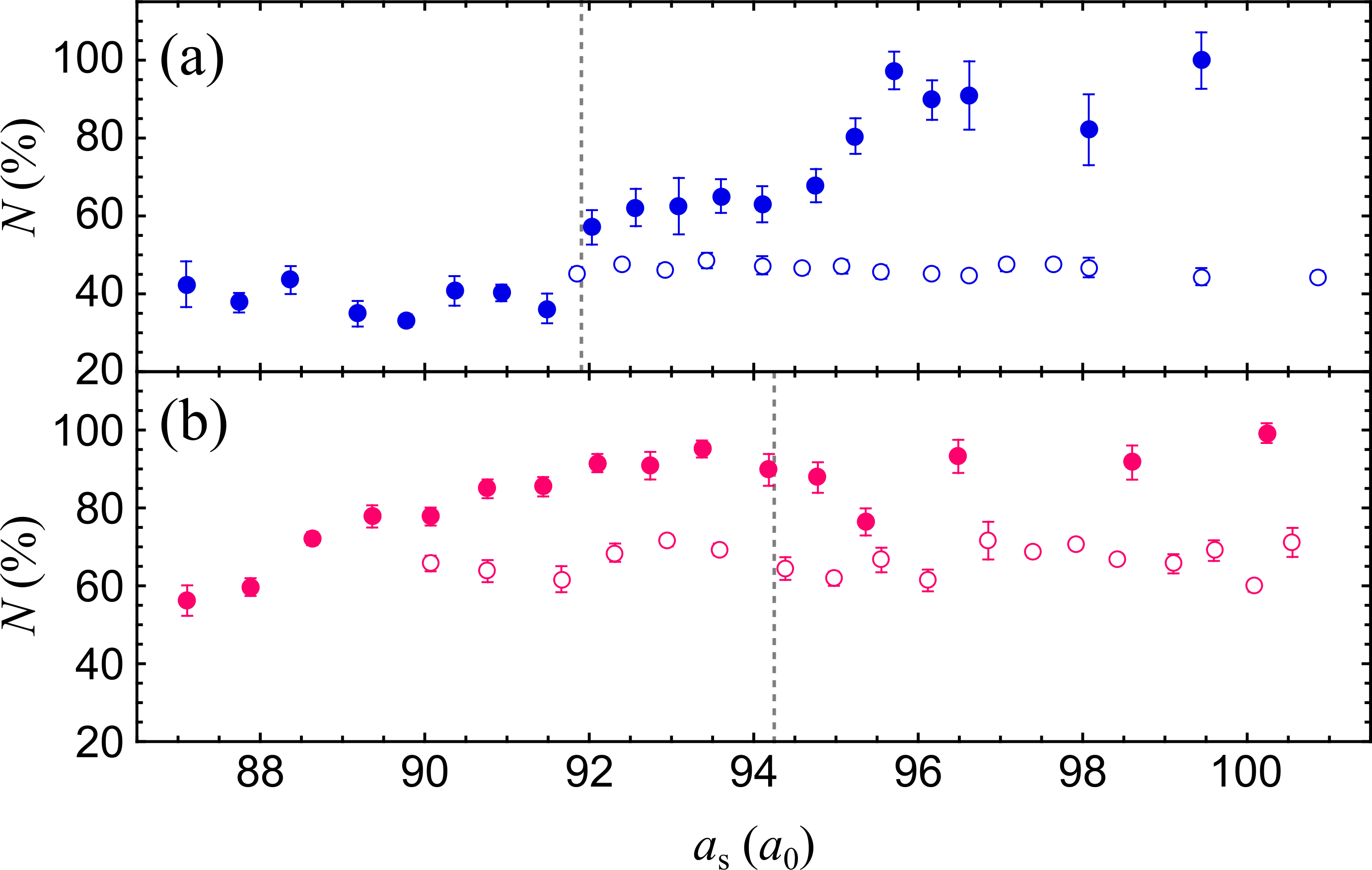}
    \caption{Evolution of the atom number during the in-going (filled symbols) and out-going ramp (open symbols) respectively for potential $V_D$ (a) and $V_C$ (b). The atom number is normalized to its maximum value on the BEC side.}
    \label{fig:expN}
\end{figure}

In the experiment, the atom number tends to decrease as one moves from the superfluid into the supersolid phase, due to three body losses, ${dN}/{dt}=-K_3\rho^2$, as depicted in Fig.~\ref{fig:expN}. The losses increase mainly because $\rho$ increases, while $K_3$ is approximately constant. In both cases, one reaches the supersolid regime with about 40\% of the initial $N$. On the way back, $N$ stays approximately constant, since the density decreased during the permanence in the supersolid regime and losses are less effective. The similar behavior of the atom number across the transitions shows that losses are similar in the two traps. The atom number in the supersolid quoted in the main text ($N=3\times 10^4$) correspond to the measured atom number just after the transitions, with an uncertainty $\delta N=\pm5\times 10^3$. We checked in the simulations that in this interval of $N$ the character of the phase transitions does not change, for both potentials.

\subsection{\label{sec:exp_kinen}Analysis of the expansion energy}
\begin{figure}
    \centering
    \includegraphics[width=0.48\textwidth]{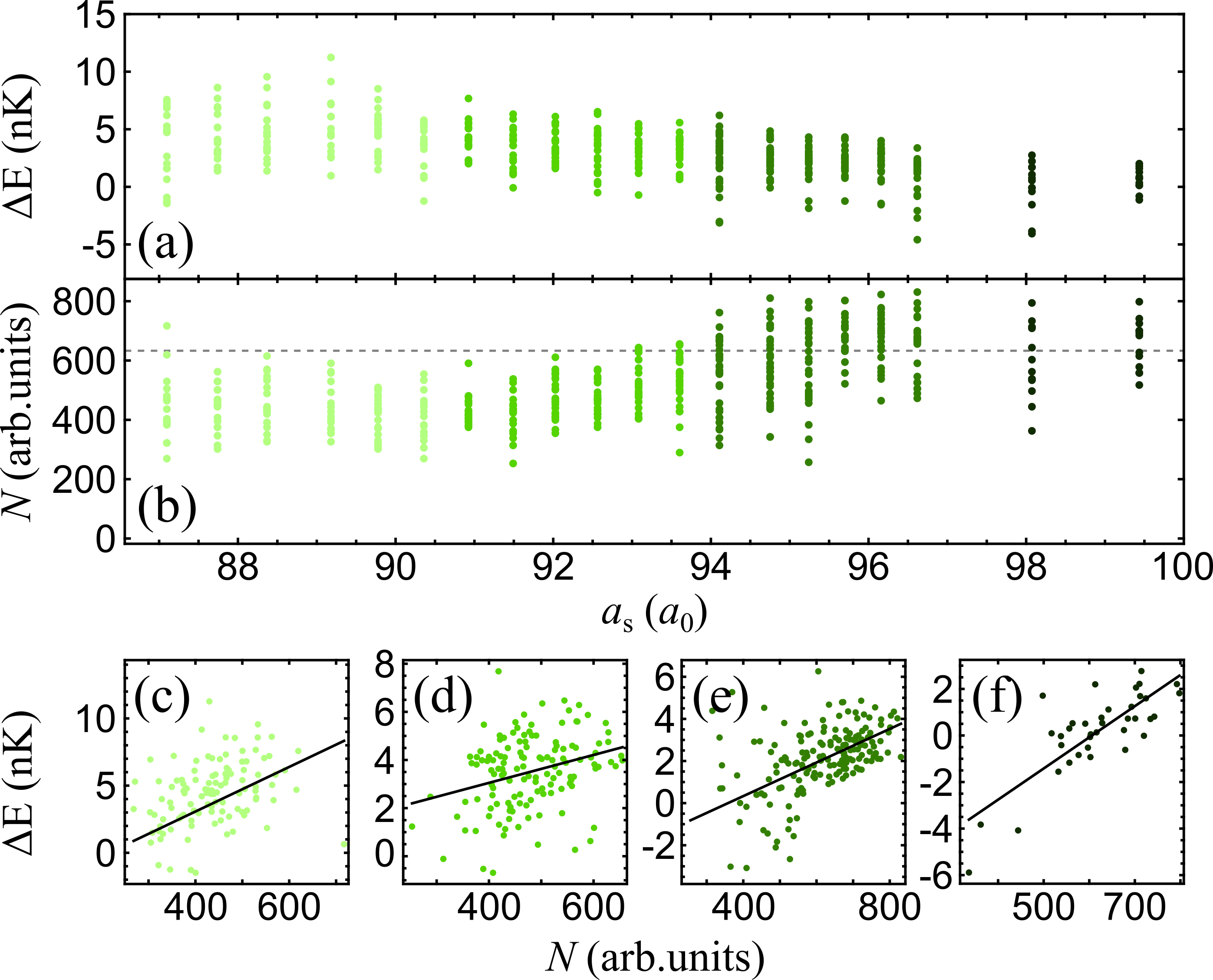}
    \caption{Analysis of the expansion energy presented in Fig.~\ref{fig:KinEn}. As an example we show the experimental data for the in-going ramp in the potential $V_D$. Panels (a-b) show excess energy (raw data) and atom number as a function of the scattering length. The horizontal dashed line marks the average atom number $\bar{N}$ in the superfluid side. Different colours correspond to the subsets in panels (c-f) where the solid lines are the linear regressions used to rescale raw data in panel (a) using Eq.~(\ref{eq:corcor}).}
    \label{fig:KinEn_analysis}
\end{figure}

We use the momentum distributions $\rho(k_x,k_y)$ also to extract the expansion energy, as shown in Fig.~\ref{fig:KinEn}. First, we compute $\delta \rho = \rho - \langle \rho \rangle$, subtracting from each distribution in the dataset the average of all the images for the corresponding harmonic potential (800 and 500 images for potentials $V_D$ and $V_C$, respectively). The excess energy $E - \langle E \rangle$ is given by
\begin{equation}
\Delta E_{raw} = \left(\frac{\hbar^2}{2m}\right) \int \delta \rho(k_x,k_y) k^2 dk_x dk_y\,.
\label{eq:Eraw}
\end{equation}
Since we use normalized momentum distributions, $\int \rho(k_x,k_y) d^2k=1$, Eq.~\eqref{eq:Eraw} gives the energy per particle. In order to eliminate the effect of atom losses, the data are corrected taking into account the correlations between energy $\Delta E_{raw}$ and the atom number, given mainly by the repulsive interactions which cannot be neglected during the expansion. As shown in Fig.~\ref{fig:KinEn_analysis}, we use a linear regression of the form $\Delta E_{\text{raw}} = \gamma N + \Delta E_0$, to determine correlations in subsets with similar $N$, corresponding to different regions in scattering length. Raw data are then rescaled using the relation
\begin{equation}
    \Delta E = \Delta E_{\text{raw}} - \gamma_i\;(N - \bar{N})\,,
    \label{eq:corcor}
\end{equation}
where $\bar{N}$ is the average atoms number in the superfluid side. In this way $\Delta E$ is increased for data with $N<\bar{N}$ and decreased in the opposite case, by an amount proportional to $\gamma_i$ for each subset. In Fig.~\eqref{fig:KinEn}, we show the mean value of data obtained from Eq.~\eqref{eq:corcor}.

\section{\label{sec:dynamics}Numerical simulations of the dynamics}

We simulated the evolution of the system when crossing the phase transitions in potentials $V_C$ and $V_D$ by solving the GP equation $i\hbar\partial_{t}\psi=\delta E/\delta\psi^{*}$, by using the FFT split-step method discussed in \cite{jackson1998}. The typical size of the numerical box is $24\mu$m per side, each discretized in $128$ points.

\begin{figure}
    \centering
    \includegraphics[width=0.48\textwidth]{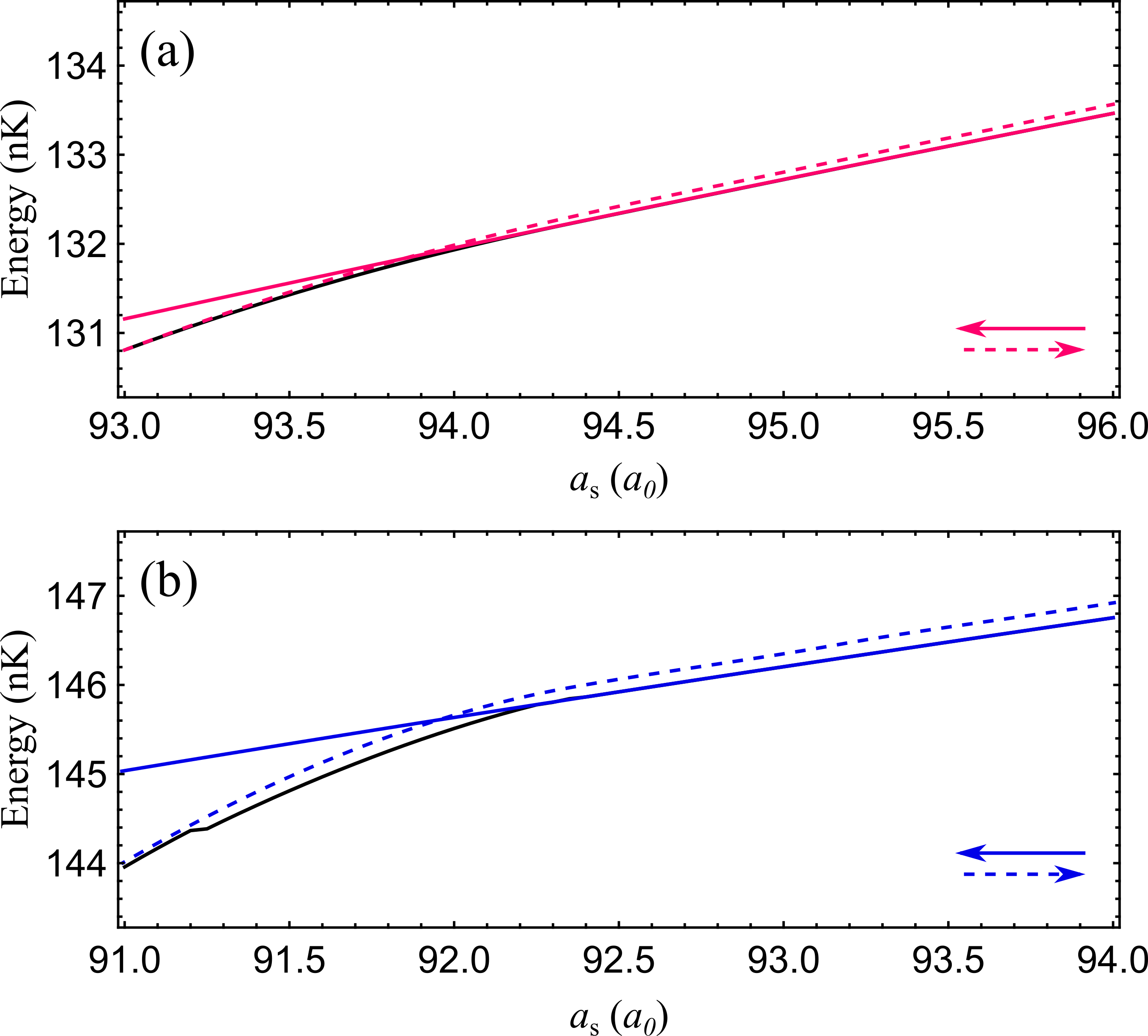}
    \caption{
    Simulated energies during a slow ramp across the two phase transitions studied in the experiment. Total energy vs scattering length for the continuous (a) and discontinuous (b) cases. Different lines are associated to the in-going ramp (solid), the out-going ramp (dashed) and the ground state energy (black).
    }
    \label{fig:simulation_energies}
\end{figure}

We found that the speed of the ramp in scattering length employed in the experiment, 0.5~$a_0$/ms, in the simulations produces effects more similar to a sudden quench of the scattering length than to an adiabatic transformation. This is presumably due to the lack of dissipation in the simulations. A ramp slower by one order of magnitude, 0.06~$a_0$/ms, results instead in a quasi-adiabatic crossing of the phase transition in potential $V_C$, see Fig.~\ref{fig:simulation_energies}, similarly to the experiment. For such a ramp speed, the simulations for
the potential $V_D$ show an increase of energy of about 1~nK, which is of the same order of magnitude of the experimental observation. A quantitative comparison of the energies in the experiment and in the simulations is not possible, mainly because of the increased contact energy in the expansion phase of the experiment. Note indeed that while the simulated energy decreases while lowering $a_s$, mainly because of the decrease of contact energy, the experimental energy in Fig. \ref{fig:KinEn} increases, because higher densities enhance the effect of the repulsive contact interaction during the expansion.

Samples of the dynamics for $V_C$ are shown in Fig. \ref{fig:DynSim} and discussed in the main text. The simulations for potential $V_D$ feature a similar behavior, i.e. excitation of collective modes of the supersolid and of the superfluid for the in-going and out-going ramps, respectively. The number of modes is however larger than for potential $V_C$, in agreement with both the larger number of modes observed in the experiment and the larger excitation energy of Fig.\ref{fig:simulation_energies}. In general, the simulations feature a larger number of modes than the experiment, presumably because they lack the dissipation that is instead present in the experiment.

We also studied the hysteresis for the discontinuous transition in potential $V_D$, in the hypothetical scenario in which the transition is crossed in both directions starting from the ground state of the system. The amplitude of the hysteresis cycle is about 0.5~$a_0$, for a holding time $t_{hold}$=40~ms. Previous simulations for a configuration close to that of potential $V_D$ found a similar result \cite{Bo19}. The hysteresis tends to disappear if one instead crosses the transition twice from the superfluid side, remaining on an excited state of the supersolid, similarly to the experimental scenario.

In conclusion, although the present simulations cannot reproduce quantitatively the observations, possibly because they do not account for temperature and dissipation effects, they support the experimental observation of a different behavior of the amplitude mode and of the energy for the continuous and discontinuous transitions, as well as the excitation of a collective mode of the superfluid following the out-going ramp.

\providecommand{\noopsort}[1]{}\providecommand{\singleletter}[1]{#1}%

\end{document}